\documentclass[manuscript,screen]{acmart}

\AtBeginDocument{%
  \providecommand\BibTeX{{%
    \normalfont B\kern-0.5em{\scshape i\kern-0.25em b}\kern-0.8em\TeX}}}

\setcopyright{acmcopyright}
\copyrightyear{2022}
\acmYear{2022}
\acmDOI{10.1145/1122445.1122456}

\acmConference[Woodstock '18]{Woodstock '18: ACM Symposium on Neural
  Gaze Detection}{June 03--05, 2018}{Woodstock, NY}
\acmBooktitle{ACM Transactions on Database Systems,
  June 03--05, 2021}
\acmPrice{15.00}
\acmISBN{978-1-4503-XXXX-X/18/06}

\usepackage{multirow}
\usepackage{mathtools}
\DeclarePairedDelimiter\ceil{\lceil}{\rceil}

\begin{document}

\title {Efficient sorting, duplicate removal, grouping, and aggregation}

\author{Thanh Do}
    \authornote{Work done at Google Inc.}
    \email{T.Do@Celonis.com}
    \affiliation{
        \institution{Celonis Inc}
        \city{Madison}
        \state{Wisconsin}
        \country{USA}
    }

\author{Goetz Graefe}
    \affiliation{
        \institution{Google Inc}
        \city{Madison}
        \state{Wisconsin}
        \country{USA}
    }
    \email{GoetzG@Google.com}

\author{Jeffrey Naughton}
    \authornote{Work done at Google Inc.}
        \affiliation{
        \institution{Celonis Inc}
        \city{Madison}
        \state{Wisconsin}
        \country{USA}
    }
    \email{J.Naughton@Celonis.com}

\begin{abstract}

Database query processing requires algorithms for duplicate removal, grouping, and aggregation. Three algorithms exist: in-stream aggregation is most efficient by far but requires sorted input; sort-based aggregation relies on external merge sort; and hash aggregation relies on an in-memory hash table plus hash partitioning to temporary storage. Cost-based query optimization chooses which algorithm to use based on several factors including the sort order of the input, input and output sizes, and the need for sorted output. For example, hash-based aggregation is ideal for output smaller than the available memory (e.g., Query~1 of TPC-H), whereas sorting the entire input and aggregating after sorting are preferable when both aggregation input and output are large and the output needs to be sorted for a subsequent operation such as a merge join.

Unfortunately, the size information required for a sound choice is often inaccurate or unavailable during query optimization, leading to sub-optimal algorithm choices. In response, this paper introduces a new algorithm for sort-based duplicate removal, grouping, and aggregation. The new algorithm always performs at least as well as both traditional hash-based and traditional sort-based algorithms. It can serve as a system’s only aggregation algorithm for unsorted inputs, thus preventing erroneous algorithm choices. Furthermore, the new algorithm produces sorted output that can speed up subsequent operations. Google’s F1~Query uses the new algorithm in production workloads that aggregate petabytes of data every day.

\end{abstract}



\maketitle

\section{Introduction}

There is a wide variety of algorithms for duplicate removal, e.g., in SQL queries like “select distinct A, B from{\dots}”. Most of these algorithms are also suitable for grouping and aggregation, e.g., in SQL queries like “select A, B, count (*), sum (X), avg (Y), min (Z){\dots} from{\dots} group by A, B”. If the data in the “from” clause are already sorted on “A, B” or something equivalent, in-stream grouping and aggregation is very simple and very efficient. If the input is sorted on a prefix of the required sort key, e.g., only on “A”, then the algorithms below apply one segment at a time, e.g., for grouping on “B” within segments defined by distinct values of “A”. Otherwise, if the output size is such that in-memory computation suffices, avoiding any need for temporary storage on external devices, then the concerns and techniques below apply to data movement between CPU caches and system memory, even if discussed here only for system memory and temporary external storage. If the input size and its storage location are such that parallel computation is desirable, partitioning permits local and independent computation of the query result, e.g., partitioning on “hash(A,B)”. If re-partitioning (shuffle, exchange) is required, best-effort in-memory duplicate removal, grouping and aggregation can reduce the communication effort. What remains is the need for an efficient sequential algorithm for duplicate removal, grouping, and aggregation of large unsorted inputs.

\begin{figure}
\centering
\includegraphics{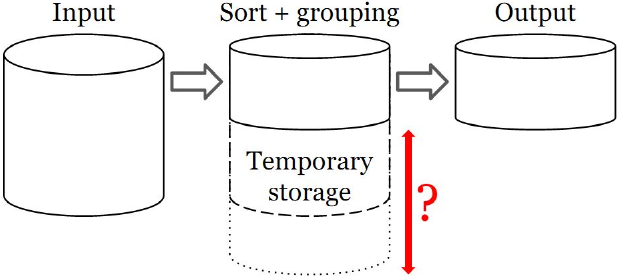}
\caption{Optimization opportunity in sorting and grouping.}
\label{figure_1}
\end{figure}

Figure~\ref{figure_1} illustrates the principal optimization opportunity in a sequential grouping algorithm for unsorted inputs. The input and output may be stored files (as shown) or they may be transient data streams. More importantly here, the sizes of input and output are fixed and their costs cannot be avoided or reduced by optimizing the grouping algorithm. The biggest optimization opportunity within the grouping operation is avoiding or reducing the need for temporary storage. If both input and output are larger than the available memory, pipeline-breaking “stop-and-go” algorithms cannot avoid temporary storage altogether. The question is whether requirements for temporary storage equal the output size, equal the input size, or exceed both sizes, e.g., due to multi-level partitioning or merging.

For unsorted inputs, there are two kinds of grouping algorithms, partitioning and merging. Both are classic divide-and-conquer designs. The first kind of algorithm hash-partitions the data into disjoint subsets, either in memory, usually as buckets in a hash table, or on temporary storage, often called partitions or overflow files. The second kind of algorithm sorts the data on all columns (fields, attributes) for duplicate removal or on the grouping columns for grouping and aggregation. The standard sort algorithm is an external merge sort with a variety of optimizations for performance and for graceful degradation, e.g., an incremental transition from in-memory sorting to external sorting. Some implementations employ a mixed approach, e.g., a hash table in memory and merge sort as external algorithm. For example, Boncz et~al.~\cite{BNE:13} mention “hash-based early aggregation in a sort-based spilling approach.” Another example for this mixed approach is the initial implementation of duplicate removal, grouping, and aggregation in Google's F1~Query~\cite{S+:18, S+:13}.

If the output is smaller than the available memory, i.e., when an in-memory hash table can accumulate the entire output, this is regarded as the algorithm of choice. TPC-H~Query~1 with four output rows for any table size (benchmark scale factor) is a prototypical example. On the other hand, if a sort order aids not only duplicate removal but also subsequent grouping or join operations, also known as interesting orderings~\cite{SAC:79}, a sort-based algorithm can reduce the total execution cost of the entire query plan. Unfortunately, due to errors in compile-time cardinality estimation, this choice is difficult and error-prone. Rather than introduce new query optimization ideas, we strive to render interesting orderings vs efficient aggregation a false choice.

Sort- and hash-based query processing are more similar than commonly recognized~\cite{G:93}. To wit, M\"{u}ller~et~al.~\cite{MSL:15} offer the insight that “hashing is in fact equivalent to sorting by hash value.” They err, however, in “hashing allows for early aggregation while sorting does not.” Perhaps they learned this erroneous understanding from~\cite{G:93} or~\cite{G:06}. One of the techniques introduced in the present paper eliminates this misunderstanding.

Sorting and duplicate removal are not new research topics, of course. For example, H\"{a}rder~\cite{H:77} summarizes that “eliminating duplicates can be achieved by a single sort” (not \textit{after} a sort). In a footnote, Bitton and DeWitt~\cite{BD:83} credit System~R (and thus H\"{a}rder) with duplicate elimination in intermediate runs. Neither of these papers explicitly mentions the similarity of algorithms for duplicate elimination, for grouping and aggregation, and for minimizing the invocation frequency of expensive operations~\cite{HN:96}, e.g., of fetching rows by row identifiers, of index searches in index nested-loops join also known as look-up join, of nested queries, and of user-defined functions.

\begin{figure}
\centering
\includegraphics{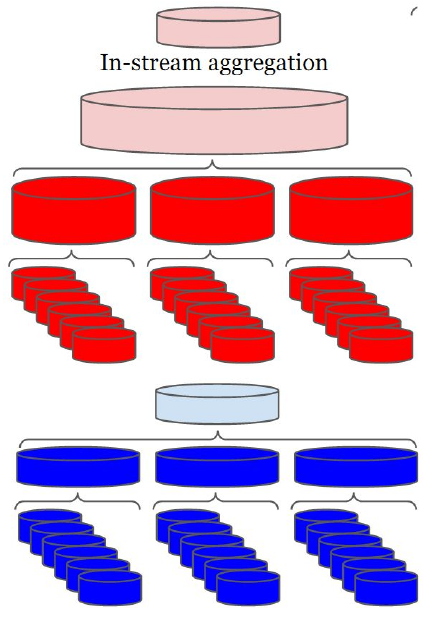}
\caption{In-stream duplicate removal after sort (red) vs. in-sort duplicate removal within runs (blue).}
\label{figure_2}
\end{figure}

Figure~\ref{figure_2} illustrates the beneficial effects of duplicate removal within runs~\cite{BD:83, H:77}. This small example assumes that input and memory sizes force 18 initial runs on temporary storage and that memory and page sizes limit the merge fan-in to 6. On the top, after a traditional external merge sort generates and merges runs, it pipelines the output of the final merge step into an in-stream aggregation operation. The result of the sort is just as large as the unsorted input; only the subsequent in-stream aggregation reduces the data volume. On the bottom, duplicate removal within runs reduces the data volume on temporary storage. Intermediate runs are never larger than the final result, which the final merge step computes. Duplicate removal reduces the output size in any merge step with the combined size of the merge inputs larger than the operation's final output.

The present paper introduces two new techniques. Both improve external merge sort in the context of duplicate removal, grouping, and aggregation; and both employ in-memory indexes where traditional designs employ priority queues. The first new technique, early aggregation, improves run generation or the input phase of external merge sort. It matches a commonly cited advantage of hash-based duplicate removal, grouping, and aggregation for unsorted input and in-memory results, e.g., for TPC-H Query~1. The second new technique, wide merging, improves the final merge step or the output phase of external merge sort. Together, these two techniques ensure that sort-based duplicate removal, grouping, and aggregation is competitive with hash-based algorithms for any input size and any output size. Of course, sort-based query processing has other advantages commonly known as interesting orderings~\cite{SAC:79}. These advantages also apply to other sort-based dataflow environments, e.g., MapReduce~\cite{DG:08} and its many successors.

A single algorithm for duplicate removal, grouping, and aggregation with robust performance (matching the best prior algorithms in all operating conditions) is more than an intellectual curiosity for the algorithm enthusiast. In many practical ways, it benefits any production system, not only in terms of code volume and effort for code maintenance but also in terms of query optimization complexity and uncertainty in algorithm choices. Other benefits apply to query execution policies, e.g., the complexity of memory management, and to physical database design, application tuning, data center monitoring, and user education.
“Generalized join”~\cite{G:11:gjoin, G:12:gjoin} represents the same intention focused on binary operations.

\begin{table}
\parbox{.45\linewidth}{
\centering
\begin{tabular}{|l|l|}
\hline
\textbf{Condition} & \textbf{Query Optimization Choice} \\
\hline
Sorted Input? & In-stream aggregation \\
\hline 
Output < Memory & \multirow{2}{*}{Hash aggregation} \\
\cline{0-0}
Unsorted output ok? & \multirow{2}{*}{}\\ 
\hline
Input/Output < fan-in & Traditional in-sort aggregation \\
\hline
Otherwise & Hash aggregation + sort \\
\hline
\end{tabular}
\caption{Traditional decision procedure.}
\label{table:complex-choice}
}
\hfill
\parbox{.45\linewidth}{
\centering
\begin{tabular}{|c|c|}
\hline
\textbf{Condition} & \textbf{Query Optimization Choice} \\
\hline
Sorted input? & In-stream aggregation \\
\hline 
Otherwise & New in-sort aggregation \\
\hline
\end{tabular}
\caption{Decision procedure with the new algorithm.}
\label{table:simple-choice}
}
\end{table}

Tables~\ref{table:complex-choice} and~\ref{table:simple-choice} illustrate this point using the decision procedures that select algorithms for duplicate removal, grouping, and aggregation. In the traditional context (Table~\ref{table:complex-choice}), one of the decisions hinges on cardinality estimation and resource availability, i.e., the relative sizes of memory size and output size. However, cardinality estimates are notoriously unreliable, in particular after prior join operations. With the proposed algorithm, the decision procedure becomes rather simple and it relies only on information about sort order (Table~\ref{table:simple-choice}), which is usually readily available during database query optimization.

In the implementation of Google’s F1~Query~\cite{S+:18, S+:13}, hash join applies recursive partitioning using a sequence of hash functions whereas hash aggregation resolves overflow by external merge sort. Adding hash partitioning to the existing in-memory hash aggregation suggests itself, but it turns out that sort-based duplicate removal, grouping, and aggregation can always be as fast – and much faster when interesting orderings~\cite{SAC:79} matter.

\begin{figure}
\centering
\includegraphics{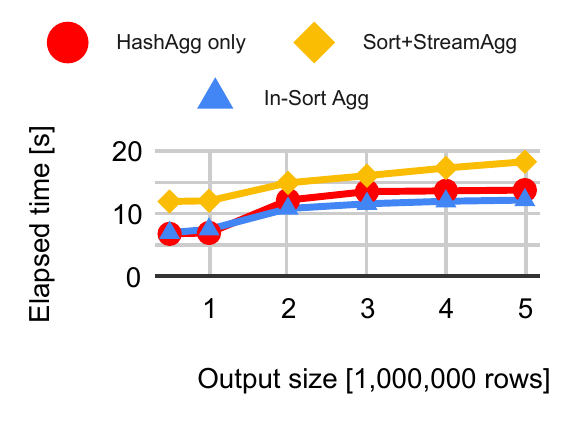}
\caption{Motivating performance example from Google’s F1~Query.}
\label{figure_3}
\end{figure}

Figure~\ref{figure_3} compares the performance of duplicate removal in F1~Query for an unsorted input of 6,000,000 rows, memory for 1,000,000 rows, and a variety of output sizes. All algorithms are implemented and tuned for production. A traditional external merge sort with subsequent in-stream aggregation is the most expensive option in all cases. Traditional hash aggregation performs very well if the output fits in memory and it degrades somewhat gracefully for outputs larger than memory, spilling an entire partition at a time. Of course it does not produce sorted output for the next operation, e.g., a join focusing on the same foreign key as the grouping operation, which is a fairly common query pattern. In-sort aggregation with the new techniques performs slightly worse than hash aggregation for small outputs and somewhat better than hash aggregation for large outputs. Given competitive performance, more graceful degradation (spilling the minimal count of rows), and the ability to produce output in interesting orderings, in-sort aggregation seems suitable as the only algorithm for duplicate removal, grouping, and aggregation for unsorted inputs.  Google’s F1~Query uses the new algorithm in production workloads that aggregate petabytes of data every day~\cite{Napa21}.

To summarize our contributions mentioned in the introduction, this paper
\begin{enumerate}
\item
introduces two techniques that speed up sort-based duplicate removal, grouping, and aggregation;
\item
analyzes and compares CPU effort and overflow I/O volumes of hash-based algorithms and improved sort-based algorithms;
\item
reports on their implementation in Google's F1~Query and on their performance; and
\item
enables competitive SQL database query processing with only a single algorithm for duplicate removal, grouping, and aggregation over unsorted inputs.
\end{enumerate}

Among the following sections, the next one reviews related prior work. Section~\ref{sec_early_agg} introduces sort-based early aggregation while consuming unsorted inputs, i.e., before in-memory sorting and thus long before writing initial runs to temporary storage~\cite{BD:83}. Section~\ref{sec_wide_merging} introduces wide merging in the final merge step, i.e., with a fan-in much higher than traditional merging. Section~\ref{sec_analysis} analyzes and compares sort- and hash-based algorithms for duplicate removal in light of these new algorithm improvements. Section~\ref{sec_impl} provides some background on our product implementation and Section~\ref{sec_perf} details performance measurements. The final section summarizes and offers a few conclusions.

\section{Related prior work}

This section reviews prior work on query processing in relational databases, in particular sorting, hashing, and grouping algorithms.

\subsection{Interesting orderings}

From early relational database management systems, sort-based algorithms and sort order have been central to query processing engines. Early research into query evaluation and grouping algorithms~\cite{E:79} discussed duplicate removal within sort operations and in-stream grouping for sorted inputs: “\dots first project the needed domains and then sort on the by-list being careful not to remove duplicates \dots Since the tuples are sorted in order of the by-list, each tuple read will have either the same by-list as the previous tuple, or it will be an entirely new by-list and there will be no more references to any previous by-lists.” The same research also considered grouping using in-memory hash indexes: “\dots the assumption that $B \geq P$ [memory size $\geq $ output size] is commonly true in practice. To the extent that this holds, the best structure to use is hash, and sorting does not help. If $B < P$ and $U$ [output row count] is large, then sorting clearly wins.”

Early research into query optimization crystallized the concept of interesting orderings and their effect on query evaluation plans~\cite{SAC:79}. Sort-based algorithms such as merge join have obvious positive interactions with sorted storage structures such as b-tree indexes as well as queries with “order by” clauses. Multiple joins on the same primary key and foreign keys are common in re-assembly of complex objects after relational normalization in the database. Grouping on foreign keys is common because it computes an aggregate property of the larger entity, e.g., the total value of all line items in a purchasing order. Thus, grouping operations before or after primary key-foreign key joins are found in many queries and query evaluation plans.

In queries with a single “select\dots from\dots where\dots group by\dots” block, the default evaluation plan executes grouping and aggregation after all joins. It has long been understood that query optimization may modify this sequence order in many queries, e.g., to reduce join input sizes~\cite {YanLarson-95-eager-lazy-aggregation-qo}. Another advantage might be pushing the grouping and aggregation logic into the sort required for the join algorithm. While an important and powerful set of optimizations, the present paper focuses on run-time efficiency and ignores these and many other query planning techniques.

\subsection{Applications of sort-based grouping and aggregation}

The algorithms discussed in this paper support sort-based duplicate removal, grouping, and aggregation. These discussions go beyond earlier descriptions of sorting and duplicate removal in relational database management systems~\cite{BD:83, E:79, G:06, H:77}. A related operation avoids redundant invocations of expensive operations such as (correlated) nested queries and user-defined functions~\cite{HN:96}.

A typical example of a large duplicate removal operation is counting distinct users in a popular web service. Logs generated by web servers may produce billions of log records per day. A dataflow pipeline or a SQL query extracts user identifiers and then removes duplicates, i.e., multiple log records pertaining to the same user. For a popular web service, this reduces many billions of rows to many millions of rows.

If counts are desired per hour or per country, the required grouping operation can use the same algorithm. In hash-based query processing, one operation (with hash table and hash-partitioning to overflow files on temporary external storage) removes duplicate user ids and another operation (with its own hash table and hash-partitioning) counts users per hour and country. In sort-based query processing, a single sort operation (on country, hour, and user identifier) serves both duplicate removal and subsequent grouping. Queries of the form “select count (distinct...)... group by...”, i.e., the combination of duplicate removal and grouping, can always benefit from interesting orderinga but should, of course, also benefit from the most efficient available algorithms for duplicate removal and grouping.

“Integrated join”~\cite{G:94, GBC:98} and “group-join”~\cite{NM:04} use a single hash table for both grouping and join. They are particularly effective when grouping and joining on the same foreign key. Due to their asymmetry, they inhibit role reversal, whereas in sort-based query processing, in-stream grouping naturally applies to both inputs of merge join as well as its output. For unsorted join inputs, the sort logic can apply duplicate removal, grouping, and aggregation.

“Generalized join”~\cite{G:11:gjoin, G:12:gjoin} includes merge logic that consumes many runs with a single buffer page, using most of memory for an in-memory index. In that sense, it is similar to wide merging (Section~\ref{sec_wide_merging}). The prior work~\cite{G:12:gjoin} stipulates an algorithm for duplicate removal, grouping, and aggregation based on generalized join with output candidate rows serving as build input and input rows as probe input, whereas wide merging does not distinguish between build and probe inputs. Nonetheless, wide merging as implemented in F1~Query may be interpreted as a first industrial implementation and deployment of “generalized distinct.” Similarly, early aggregation (Section~\ref{sec_early_agg}) may be seen as an implementation of the input phase or of a hybrid phase of generalized distinct.
Due to a lack of understanding and appreciation of offset-value coding, those earlier attempts might achieve competitive I/O volumes, i.e., overflow partitions in hash join and runs in external merge sort, but they do not achieve equivalent column value accesses and thus CPU effort (Section~\ref{sec_column_accesses}).

Rollup functionality has existed for a long time in programming environments such as Cobol and been suggested for database queries~\cite{G:93}. Sort-based aggregation can compute multiple levels of aggregation with a single sort operation, e.g., for a query of the form “select \dots group by rollup (year, month, day)”. In contrast, hash-based aggregation requires separate computations for each level of aggregation. Each level requires a hash table and possibly partitioning to temporary storage.

Log-structured merge-forests and stepped-merge forests~\cite{JNS:97, OCG:96} are nearly ubiquitous in key-value stores. In this context, runs are often called deltas and merging is often called compaction, because merging includes aspects of aggregation and compression. The individual merge steps are similar to those of external merge sort, but their merge policies (what to merge when) are quite different for multiple reasons. First, their input is assumed endless. For example, it is not possible to delay merging until run generation is complete; merging must be concurrent to run generation. Second, inputs include traditional insertions, which are mapped to append operations, as well as updates, which are mapped to insertions of replacement rows, and deletions, which are mapped to insertions of “tombstone” rows. The merge logic aggregates insertions, updates, and deletions either into a final state or into a history of versions, including removal of out-of-date versions (also known as garbage collection). Third, individual runs are formatted as b-trees, not flat files, in order to permit search and queries over recent as well as historical information. Alternative formats include a single partitioned b-tree, with runs mapped to partitions. Bit vector filters can enable a query to skip some partitions and thus improve performance. Fourth, the merge fan-in and the frequency of merge steps are controlled not by the memory size but by the desire for good query performance, i.e., searching few partitions. Many designs and deployments of log-structured merge-forests employ low-fan-in merge steps, even binary merging.

Decades ago, Gray suggested sorting recovery log records on the database page identifier to which they pertain~\cite{G:78}: “This compressed redo log is called a change accumulation log. Since it is sorted by physical address, media recovery becomes a merge of the image dump of the object and its change accumulation tape.” It seems a small step from sorting recovery log records to building b-tree indexes, another step to building indexes incrementally and continuously (in the manner of log-structured merge-forests), and yet another small step to using such indexes for page-by-page incremental, on-demand, seemingly instant recovery from single-page failures, from system failures (software crashes), and from media failures~\cite{GGS:16}.

\subsection{Optimizing “group by” and “order by”}

Functional dependencies enable many interesting optimizations for “group by” and “order by” queries~\cite{SSM:96}. Functional dependencies are implied by primary key integrity constraints and by prior “group by” operations.

More specifically, a “group by” clause requires a set of columns (expressions) and an “order by” clause requires a list of columns. Functionally dependent columns can be removed anywhere in a set but only in subsequent positions within a list. For example, in two database tables for purchase orders and their line items, with o\_orderdate functionally dependent on o\_orderkey, the first three among the following four clauses permit simplification but the last one does not:

\begin{enumerate}
\item “{\dots}group by o\_orderkey, o\_orderdate”,
\item “{\dots}order by o\_orderkey, o\_orderdate”,
\item “{\dots}group by o\_orderdate, o\_orderkey”,
\item “{\dots}order by o\_orderdate, o\_orderkey”.
\end{enumerate}

Below is a (first) example of using functional dependencies in an unusual way. The first query seems to require grouping and aggregation after the join, but the second and third queries are essentially equivalent to the first one due to the functional dependency of order date on the grouping key. Adding a functionally dependent column to a “group by” clause applies the insights of~\cite{SSM:96} in the reverse direction. As grouping key and join key are the same, order date is a constant within each group of line items. The fourth query variant is equivalent to the first query and most conducive to efficient execution. Note that the many-to-one join changes into a one-to-one join.

{\small
\begin{enumerate}
\item \begin{verbatim}
select o_orderkey, avg (l_shipdate - o_orderdate)
from orders, lineitem
where o_orderkey = l_orderkey
group by o_orderkey
\end{verbatim}
\item \begin{verbatim}
select o_orderkey, o_orderdate, 
    avg (l_shipdate - o_orderdate) 
from orders, lineitem
where o_orderkey = l_orderkey
group by o_orderkey, o_orderdate
\end{verbatim}
\item \begin{verbatim}
select o_orderkey, o_orderdate, 
    avg (l_shipdate) - o_orderdate
from orders, lineitem
where o_orderkey = l_orderkey)
group by o_orderkey, o_orderdate
\end{verbatim}
\item \begin{verbatim}
select o_orderkey, avg_shipdate - o_orderdate
from orders,
    (select l_orderkey, 
        avg (l_shipdate) as avg_shipdate
     from lineitem
     group by l_orderkey) as a 
where o_orderkey = l_orderkey
\end{verbatim}
\end{enumerate}
} 

In many queries, query rewriting such as this example is required to enable integrated join or group-join. In integrated join, optimizing grouping and join on the same column (set) applies only to the build input; in group-join, it applies only to the probe input. In sort-based query processing, grouping and join on the same column (set) enjoy the benefits of interesting orderings if grouping is applied to either of the two join inputs, or even to the join output. In other words, interesting orderings benefit query performance whether or not query optimization applies all kinds of clever and uncommon rewrites.

In sum, sort-based duplicate removal, grouping, and aggregation can benefit from proper use of functional dependencies because they permit optimizations of both grouping and ordering, but it seems that sort-based query evaluation plans are somewhat more forgiving and flexible than hash-based query execution.

\subsection{Optimizations of sort operations}


High-performance sorting requires efficiency, scalability, and robustness of performance. Efficiency may benefit from tree-of-losers priority queues, normalized keys, offset-value coding, and hardware support.
Among the techniques mentioned, normalized keys encode key values in order-preserving strings such that all subsequent key comparisons use intrinsics or hardware instructions for binary strings. Further hardware support may focus on tree-of-losers priority queues and offset-value coding in normalized keys, e.g., the UPT “update tree” instruction and the CFC “compare and form codeword” instruction~\cite {I:05}. Iyer~\cite {I:05} sums up that “Together, the UPT and CFC instructions do the bulk of sorting in IBM’s commercial DBMS DB2 running on z/Architecture processors.”

Offset-value coding~\cite {C:77} encodes one row’s key value relative to another row’s key value that is earlier in the sort sequence. Offset and value are combined into an integer such that a single machine instruction can compare two offset-value codes and decide a comparison of two rows encoded relative to the same base row. Offset-value codes can decide many comparisons in tree-of-losers priority queues.

Table~\ref {table:ovc} illustrates the derivation of descending and ascending offset-value codes in a stream of rows in ascending sort order on all columns. With four sort columns, the arity of the sort key is 4; the example assumes that the domain of each column is 0…99. Descending offset-value codes take the actual offset but the negative of the column value. Ascending offset-value codes take the negative offset but the actual column value. Table~\ref {table:ovc} ignores that small key domains permit encoding multiple key columns together.

\begin{table*}[t]
\caption{Offset-value codes in a sorted file or stream.}
\label{table:ovc}
\centering
\begin{tabular}{|r|r|r|r|r|r|r|r|r|r|}
\hline
\multicolumn{4}{|c|}{Rows and their} &
\multicolumn{3}{c|}{Descending OVC} &
\multicolumn{3}{c|}{Ascending OVC} \\
\cline{5-10}
\multicolumn{4}{|c|}{column values} &
$offset$ & $domain - value $ & OVC & $arity - offset$ & $value$ & OVC \\
\hline
\textbf{5} & 7 & 3 & 9 & 0 & 95 & 95 & 4 & 5 & 405 \\
\hline
5 & 7 & 3 & \textbf{12} & 3 & 88 & 388 & 1 & 12 & 112 \\
\hline
5 & \textbf{8} & 4 & 6 & 1 & 92 & 192 & 3 & 8 & 308 \\
\hline
5 & \textbf{9} & 2 & 7 & 1 & 91 & 191 & 3 & 9 & 309 \\
\hline
5 & 9 & 2 & 7 & 4 & - & 400 & 0 & - & 0 \\
\hline
5 & 9 & \textbf{3} & 4 & 2 & 97 & 297 & 2 & 3 & 203 \\
\hline
5 & 9 & 3 & \textbf{7} & 3 & 93 & 393 & 1 & 7 & 107 \\
\hline
\end{tabular}
\end{table*}

Finally, a tree-of-losers priority queue~\cite {G:63, K:98}, also known as a tournament tree, embeds a balanced binary tree in an array, with the tree’s unary root in array slot 0. It is efficient due to leaf-to-root passes with one comparison per tree level; root-to-leaf passes with two comparisons per tree level are not required. Run generation and merging with tree-of-losers priority queues can guarantee sort operations with near-optimal comparison counts.

Scalability is principally about parallelism – twice the resources should process the same data in half the time (also known as speed-up) or twice the data in the same time (scale-up). Robustness of performance is about performance cliffs and graceful degradation – for example, the transition from an in-memory quicksort to an external merge sort should be gradual rather than a binary switch, such that a single additional byte or input row cannot force spilling the entire memory contents. The techniques introduced in Sections 3 and 4 are orthogonal to both scalability and robustness of performance: the new techniques do not offer improvements in those directions but they also do not impede or hinder existing or future techniques for scalability and for graceful degradation.

\subsection{Early results in join-by-grouping}

Complementing optimizations of sort-based grouping, there is a join algorithm based on grouping. It requires that the implementation of external merge sort can interleave multiple record types within memory and within each run on temporary storage. Sorting a mixed stream of records on join key values produces mixed “value packets”~\cite{K:80}, i.e., sets of rows with equal sort keys. In the context here, equal sort keys means equal join keys. Forming or combining value packets is a kind of aggregation. The join output is computed from the final sorted stream by computing a cross product within each mixed value packet. Alternatively, when the sort and merge logic forms or combines mixed value packets, it can produce join results as an immediate side effect. In other words, early aggregation in this context enables early and incremental join results. Variants of this algorithm can compute semi-join, anti-semi-join, all forms of outer join, set and bag intersection and difference (e.g., “intersect all” in SQL). Anti-semi-join and equivalent result rows of outer joins cannot be produced early.

\begin{figure*}
\centering
\includegraphics[width=10cm, height=4cm, keepaspectratio]{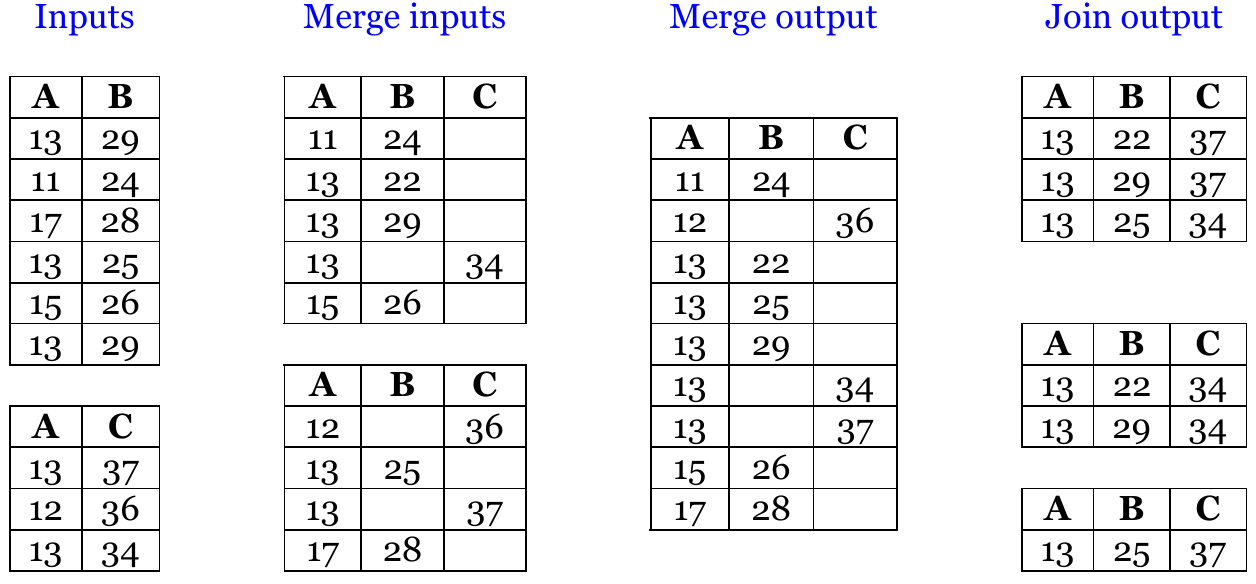}
\caption{Join by grouping.}
\label{figure_4}
\end{figure*}

Figure~\ref{figure_4} illustrates one merge step within this algorithm applied to an inner join. Two unsorted input tables (far left top and bottom) are joined on column $A$. Rows from both inputs are scanned concurrently and run generation creates initial sorted runs (center left). These runs contain multiple record types, one for each join input. Merging runs (from center left to center right) relies on the standard merge logic known from external merge sort. As a side effect, this merge step combines value packets (in this example, for $A = 13$) and produces new join results (far right top). Before this merge step, when run generation assembled the merge inputs (center left), it produced early join results (far right middle and bottom) while forming the value packets in these runs.

Once two records have been joined, they remain in the same value packet until the sort finishes. Hence, there is no danger of duplicate (redundant, wrong) output. For example, in Figure~\ref{figure_4}, the two original inputs (far left) contain 3 and 2 rows with key value $A = 13$ for $3 \times 2 = 6$ rows in the join result; the three partial results (far right top to bottom) contain precisely these 6 rows. When the sort finishes, the final value packets are dropped; the operation’s output is the join result computed incrementally as side effect of forming and combining mixed value packets.

This join algorithm is an alternative to more complex sort-based join algorithms with early output~\cite{DST:03}. Its output rate and memory requirements mirror those of symmetric hash join~\cite{WA:91} if early aggregation and wide merging are enabled, which are the topics of the next two sections.

\subsection{Summary of related work}

To summarize our observations on related prior work, duplicate removal, grouping, and aggregation occur in a large variety of contexts, from data warehouse queries and business intelligence to analysis of web logs. Substantial research and development effort have been invested in both query optimization and query execution specifically for duplicate removal, grouping, and aggregation. A remaining thorny problem is that traditional sort- and hash-based algorithms are optimal in different circumstances (relative sizes of input, output, and memory; sort orders interesting for subsequent operations), rendering a choice during compile-time query optimization difficult and error-prone. Instead, the next two sections offer a single algorithm that, assuming equal care in algorithm implementation, always matches the best traditional algorithm for duplicate removal, grouping, and aggregation, at least in terms of data movement (including I/O) and of column accesses and comparisons.

\section{Early aggregation during run generation}
\label{sec_early_agg}

Techniques for early duplicate removal, grouping, and aggregation are particularly valuable for queries with small results, i.e., duplicate removal or aggregation with many input rows and few output rows. More specifically, if the output fits in the memory allocation available for the grouping operation but the input is unsorted and large (such that expensive spilling to temporary storage is required in a traditional sort algorithm), then early aggregation improves the performance of sort-based aggregation. In fact, early aggregation ensures that sort-based aggregation spills no more data to temporary storage than hash-based aggregation and sometimes a little bit less.

Early aggregation pertains to the input phase of an external merge sort, i.e., run generation. Traditional run generation employs read-sort-write cycles or replacement selection. The former uses an internal sort algorithm such as quicksort for initial runs as large as memory; the latter uses a priority queue and can produce initial runs twice as large as memory for truly random input, as large as memory in the worst case, and as large as the entire input in the very best case.

\begin{figure}
\centering
\includegraphics{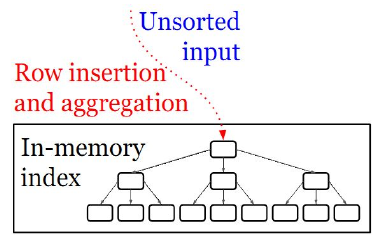}
\caption{In-memory aggregation.}
\label{figure_6}
\end{figure}

In contrast, early aggregation eschews both quicksort and priority queues; instead, it uses an ordered in-memory index, e.g., an in-memory b-tree. Such an index enables both read-sort-write cycles and replacement selection. More importantly, an index enables immediate discovery of duplicate key values, just like a hash table. If the output size is smaller than the memory size, early aggregation avoids all I/O to spill intermediate data to temporary storage. Figure~\ref{figure_6} illustrates this case.

\begin{figure}
\centering
\includegraphics{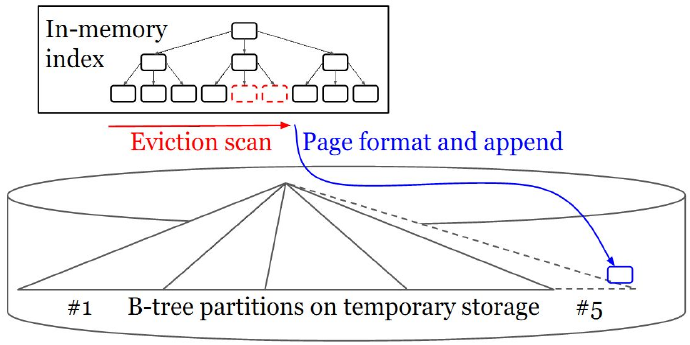}
\caption{Run generation using an ordered in-memory index.}
\label{figure_5}
\end{figure}

Figure~\ref{figure_5} illustrates run generation using an ordered in-memory index. Due to insertions of rows and key values from the unsorted input, the index grows continuously as new key values create new index entries. However, key values equal to ones already in the index are absorbed by aggregation. In the ideal case, the entire input can be absorbed within memory. Otherwise, a scan thread evicts keys and rows from leaf pages, either occasionally in read-sort-write cycles or continuously in replacement selection, and writes them to runs on temporary storage, where they form a forest (of many trees) or a partitioned b-tree (of many partitions). Early aggregation does not prescribe a choice between read-sort-write cycles and replacement selection - this choice is left to the implementer's preference, e.g., for simple space management even with variable-sized rows, or to considerations beyond the discussion here, e.g., techniques for  memory-adaptive query execution~\cite {PangCareyLivny-93-Sorting}.

\subsection{Example 1}
\label{sec_early_agg_ex1}

As an archetypal example of a grouping query with a small output, consider Query~1 of the TPC-H benchmark~\cite{TPC}. In scale factor $SF = 1$ of the benchmark, the query scans a database table of about $6M$ rows, selects about $90\%$ of those, and then reduces them to four rows with counts, sums, and averages. For scale factor $SF =$ 1,000, the query reduces $6B$ rows to four output rows. Clearly, an external merge sort of $6M$ or even $6B$ rows with subsequent grouping and aggregation, i.e., the two techniques of Figure~\ref{figure_2}, cannot compete with an algorithm that uses a hash table to simply accumulate the final query result within a small memory allocation of 4 rows.

There are many real-world queries in which a grouping result fits within the readily available memory allocation. For example, if each operator within a query evaluation plan is allotted 100~MB by default, then any grouping operation with a result smaller than 100~MB should remain an in-memory operation and algorithm. This is, of course, true for hash aggregation but it can also be true for sort-based duplicate removal, grouping, and aggregation.

Figure~\ref{figure_6} illustrates this case. The in-memory index can grow until it fills memory. Skew in the key value distribution does not matter as an ordered index adapts to the actual distribution. Only if the output size exceeds the available memory, spilling to runs on temporary storage as shown in Figure~\ref{figure_5} is required.

\subsection{Example 2}
\label{sec_early_agg_ex2}

As another example, imagine the “group by” clause of Example 1 extended such that the final output is larger than memory, i.e., $O > M$ or more specifically $O = 2M$. Even with early duplicate removal, grouping, and aggregation, this example requires runs on temporary storage. As key values in the in-memory index are unique, the in-memory index immediately matches and absorbs a fraction of the input rows. With run generation by replacement selection and memory always full, about $M/O = \frac{1}{2}$ of all input rows are absorbed immediately. Ignoring the effects of an in-memory run for graceful degradation from in-memory sorting to external merge sort, the total size of all initial runs is about $M + (1 - M/O) \times I$ (for input size $I$) or in the specific example $M + \frac{1}{2}I$. With only unique key values in the in-memory index, the traditional logic for duplicate removal, grouping, and aggregation~\cite{BD:83} while writing runs on temporary storage is not required.

\begin{figure}
\centering
\includegraphics{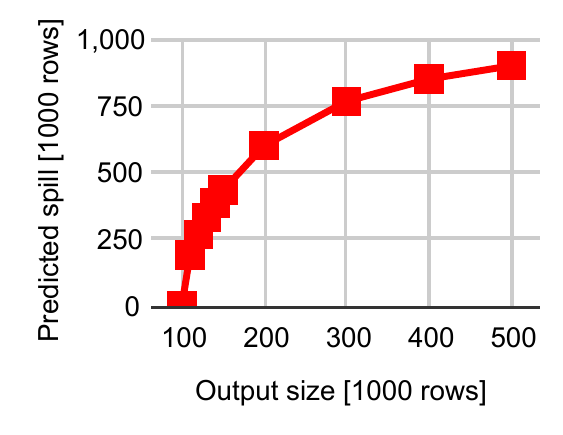}
\caption{Predicted spill volume.}
\vspace{-10pt}
\label{figure_7}
\end{figure}

Figure~\ref{figure_7} shows the predicted spill volume for input size $I =$ 1,000,000 rows and memory size $M =$ 100,000 rows.  For the left-most point, the output size equals the memory size ($O=M$) and no spilling is required. If the output is many times larger than the available memory allocation, practically all input rows spill. The calculation in this prediction assumes replacement selection using an in-memory index, even though our implementation uses read-sort-write cycles. (Recall that replacement selection is usually implemented using a priority queue and for random input produces runs twice the size of memory, whereas read-sort-write cycles are usually implemented using quicksort and produce runs the size of memory; recall also that an ordered in-memory index permits either read-sort-write cycles or replacement selection, according to the implementer's preference.)

\subsection{Algorithms and data structures}
\label{sec_early_agg_alg}

One design combining in-memory run generation with early aggregation uses two data structures. For example, Boncz et~al.~\cite{BNE:13} mention “hash-based early aggregation in a sort-based spilling approach.” If the in-memory hash table fails to absorb (i.e., to aggregate) an input row, the row is added to both in-memory data structures, i.e., the hash table as well as an array for quicksort or a priority queue for replacement selection.

An alternative design employs a single data structure for two purposes, searching and sorting. A suitable data structure is an ordered in-memory index, e.g., an in-memory b-tree~\cite{BM:72, G:11}. Note that comparisons are required only during the search. If no match is found, the search for a match has already found the insertion point as a side effect.

If the search employs a binary search within each tree node, the count of comparisons per input row is equal to that in a tree-of-losers priority queue, which is 10-20\% lower than the count of comparisons in quicksort and very close to the theoretical minimum. (Sorting $ N $ items is equivalent to selecting one of $ N! $ permutations, which requires $ \log_2 (N!) $ comparisons.)

Interpolation search within each tree node is even more efficient if the key value distribution is nearly uniform, which is likely the case if the sort key is a hash value. Note that sorting on hash values permits exploiting interesting orderings if other algorithms and storage structures also sort on hash values. Merge joins and b-trees on hash values are attractive for database columns with no real-world “$<$” comparison, e.g., practically all artificial identifiers and thus many primary keys and foreign keys in real-world databases.

The row format within the in-memory index is the same as in runs on temporary storage. It may differ from the row formats in both the input and the output. For example, in addition to a grouping key, input rows may contain a value, intermediate rows a sum and a count, and output rows an average. Similar considerations apply when computing variance, standard deviation, co-variance, correlation, regression slope and intercept, etc.

In traditional run generation, read-sort-write cycles may use quicksort to produce runs the size of memory. Run generation by replacement selection using a priority queue can produce runs the size of memory or, with an additional comparison for each new input row as well as a flag within each row in memory, twice the size of memory. Run generation using an in-memory index can produce runs twice the size of memory without an additional comparison and without a flag in each row in memory. Eviction from memory to temporary storage repeatedly scans the in-memory index as shown in Figure~\ref{figure_5}. Advancing the scan evicts rows and frees index nodes whenever the in-memory index needs to split a node and thus allocate a free node.

\subsection{Analysis}
\label{sec_early_agg_analysis}

Three questions suggest themselves for analysis:
\begin{enumerate}
\item How many comparisons are required in early aggregation, i.e., run generation with an in-memory index? How does that compare to run generation with read-sort-write cycles, e.g., quicksort, and with replacement selection, i.e., a tree-of-losers priority queue?
\item How many column accesses are required to enable these comparisons?
\item If the output size is a small multiple of the memory size, what fraction of input rows are absorbed immediately in the in-memory index, and how many rows are spilled to runs on temporary storage?
\end{enumerate}

In run generation using an efficient tree-of-losers priority queue, the count of comparisons per input record is $\log_2 (M/R)$ for memory size $M$ and record size $R$. This is correct for run size equal to memory size $M$; one additional comparison is required for expected run size $2M$. Using quicksort, the expected count of comparisons is 10-25\% higher; the worst case for quicksort is much higher. In run generation using an in-memory index with binary search, the count of comparisons per input record is again $\log_2 (M/R)$. Using interpolation search or batches of search keys, it can be substantially lower.

If the final output fits in memory, i.e., $O \leq M$, then the count of comparisons per input record is $\log_2 (O/R)$. Again, it can be substantially lower with interpolation search or batches of search keys. Linear interpolation is effective if the key value distribution in the output is practically uniform. This is probable if the keys are hash values, i.e., when sorting and grouping on hash values, at least as the leading sort key.

In a striking similarity, hash aggregation requires a search in the hash table for each input record, i.e., a hash calculation plus a scan through a hash bucket. Those are comparable to the interpolation calculation and the local search near the interpolated position. If memory remains full all the time during run generation, then each input row has a probability of $M/O$ (memory size over output size) of finding a matching key value in memory and of being absorbed in the index without insertion. If $M \geq O$, this probability is a certainty and spilling to runs on temporary storage is not required. If this probability is very small, then practically all input rows spill into runs on temporary storage.

While the count of comparisons is a traditional metric of algorithm complexity, it is neither the only metric nor the most meaningful one in practice. For example, in a table with 10 or 100 columns, hash-based duplicate removal must locate, access, decompress, and hash every column in every row in order to compute hash values, whereas sort-based duplicate removal might never access more than the first few columns if those are sufficient to decide comparisons. Section~\ref{sec_column_accesses} analyses these effects in more detail.

\subsection{Summary of early aggregation}

To summarize, early aggregation uses an in-memory index to match and absorb input rows during run generation in duplicate removal, grouping, and aggregation. In the ideal case, it entirely avoids spilling rows to temporary storage. A typical example is TPC-H Query~1 with only 4 output rows even for very large databases and input tables.

The in-memory index can be a simple b-tree or it can be optimized in many ways. A binary search guarantees $\log_2 (M)$ comparisons per input row. Replacement selection can produce runs twice as large as memory. An alternative design combines an in-memory hash table (for immediate detection of duplicates) with a priority queue (for replacement selection). This alternative suffers from twice the space overhead and twice the maintenance effort as well as the complexity of coordinated incremental insertions and deletion of rows, e.g., during graceful degradation from an in-memory operation to an external algorithm spilling to temporary storage.

Whatever the algorithm for run generation, runs require merging with the traditional merge logic known from external merge sort or, in many cases, an optimized merge logic to be known as wide merging. Interestingly, both early aggregation and wide merging replace priority queues, the traditional data structure of choice for sorting, with ordered in-memory indexes, e.g., in-memory b-trees.

\begin{figure}
\centering
\includegraphics{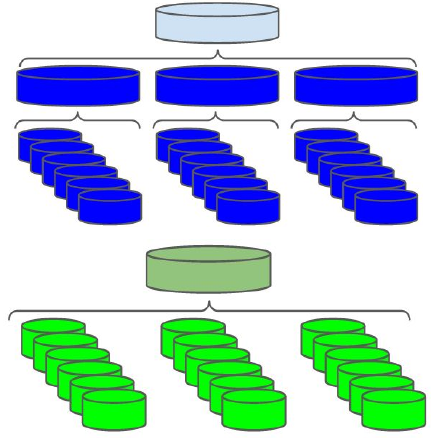}
\caption{The effect of wide merging (green).}
\label{figure_8}
\end{figure}

\section{Wide merging in the final merge step}
\label{sec_wide_merging}

For complete performance parity with hash aggregation, in-sort grouping and duplicate removal requires a final merge step with a merge fan-in potentially higher than a traditional merge step with the same memory allocation and page size (unit of I/O).

In contrast to traditional merging in an external merge sort, wide merging is not limited to a specific fan-in. Traditional merging uses its memory allocation for page buffers for the merge inputs. A page for each input limits the fan-in to the quotient of memory size and page size. Wide merging instead uses a single buffer page for all its inputs (or maybe two to implement double-buffering). The central in-memory data structure is an index mapping grouping columns to candidate output rows, similar to the index used in Section~\ref{sec_early_agg} for early aggregation. Processing one page at a time from the runs containing aggregation input, the algorithm accumulates input rows into the index. When processing many runs in a single merge step, wide merging can be much more efficient than traditional aggregation within sort, e.g., saving entire intermediate merge levels.

Figure~\ref{figure_8} extends the illustrations of Figure~\ref{figure_2} with a third merge strategy for sort-based duplicate removal, grouping, and aggregation from unsorted input data. Wide merging can consume and aggregate many more runs than it has memory pages, e.g., 18 runs with merely 6 memory pages. This is possible if memory can hold and index a key range equal to the key range of one page in the merge inputs on temporary storage. Wide merging uses only one input buffer shared by all runs. After reading a page, wide merging absorbs the page contents into its in-memory index before reading the next page, typically from a different input run.

Figure~\ref{figure_9} illustrates the flow of data in wide merging. Using a priority queue, the algorithm chooses an input run from which to read the next page. Once that page is in the input buffer, the key values on that page are found in the in-memory index. If a key value exists, the input row is absorbed without growing the index. If the key value does not exist, a new entry is required in the index. In this way, the ordered in-memory index absorbs all rows and key values from all runs on the temporary storage device. As the merge logic progresses through the domain of key values, the active key range in the in-memory index turns over continuously. The left edge of the in-memory index produces final output and the right edge adds new key values from the merge inputs.

A priority queue guides the page consumption sequence during wide merging. It has an entry for each of its input runs, e.g., 18 entries in the example of Figure~\ref{figure_8}. It is similar to the priority queue used in traditional external merge sort for guiding read-ahead within the merge input, also known as forecasting~\cite{K:98}. From each of the input runs, it tracks the highest key value already read; the next read operation targets the run with the lowest of these key values. If implemented as a tree-of-losers priority queue, one leaf-to-root traversal in the priority queue is required for each page in the runs on temporary storage, or $\log_2 (18) \approx 4$ comparisons in the example of Figure~\ref{figure_8}. The highest key value in the first page of each run initializes this priority queue. Alternatively, if the sort operation retains minimum and maximum key values of each run, e.g., for the purpose of concatenating runs with disjoint key ranges~\cite{H:77, G:06}, the retained minimum key values initialize this priority queue.

\subsection{Example 3}
\label{sec_wide_merging_ex3}

\begin{figure}
\centering
\includegraphics{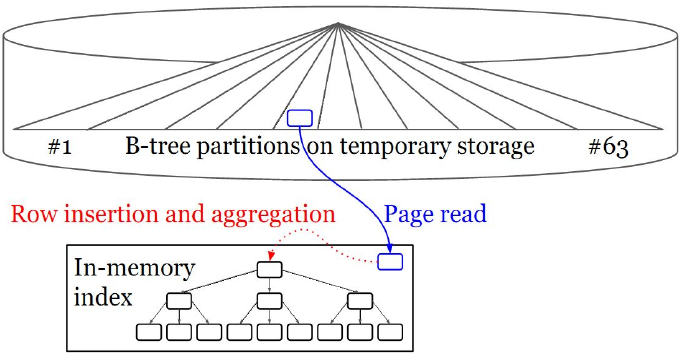}
\caption{Wide merging using an ordered in-memory index.}
\label{figure_9}
\end{figure}

Consider a specific example of wide merging and its benefits: single-threaded duplicate removal with input size $I = 750,000$ rows, memory size $M = 1,000$ rows, partitioning fan-out and merge fan-in (in traditional merge logic) and partitioning fan-out $F=6$, and final output size $O = 32,000$ rows. Importantly, the memory is much smaller than the final output and the final output is much smaller than the original input, or $M\ll O\ll I$.

In this example, hash aggregation invokes $L=2$ partitioning levels to divide all input rows into $F^2 = 36$ partitions of about $I / F^2 =  21,000$ rows each. During these partitioning steps, the output buffers are too small to enable much early (opportunistic) duplicate removal. After two partitioning levels, each partition contributes about $O / F^2 $ rows to the final output. As the output per partition is smaller than memory ($O / F^2 = 900 < 1,000 = M$), duplicate removal can occur in memory in spite of input partitions much larger than memory ($I / F^2 = 21,000 > 1,000 = M$). The total size of all temporary partitions in both partitioning levels is $2I = 1,500,000$ rows, each written and read once. More generally, hash aggregation can aggregate the remaining partitions in memory after $L \geq \log_F (O/M)$ partitioning levels, which is here $\log_6 (32) = 2$ partitioning levels.

In contrast, in-sort duplicate removal, grouping, and aggregation starts with run generation by replacement selection. Each run holds about $2M = 2,000$ rows; thus, this example requires about $I / (2M) = 376$ runs. The first merge level produces $376/F = 63$ runs of about $F \times 2,000 = 12,000$ rows. The second merge level produces $63/F = 11$ runs. Aggregation within runs~\cite{BD:83} cuts their size from $F \times 12,000 = 72,000$ rows to $O = 32,000$ rows. The penultimate merge step combines 6 of these 11 runs into another run of $O = 32,000$ rows and the last merge step produces the final output. The total size of all runs spilled to temporary storage during run generation, full merge, optimized merge, and partial merge is $750,000 + 750,000 + 11 \times 32,000 + 32,000 = 1,884,000$ rows, each written and read once. This is about 25\% more than the temporary partitions in hash aggregation.

Wide merging enables further savings. In this example, a single final merge step can aggregate the 63 runs after the first merge level. Merging 63 runs with memory for only a few input buffers requires an in-memory index for immediate duplicate removal.

A traditional merge step merging 6 of these 63 runs has its output cut to $O = 32,000$ rows. This is true whether the merge logic uses traditional single-page buffers and a priority queue or an in-memory index for wide merging. If such an index holds practically all distinct key values over the course of the merge step, it can (with the appropriate pacing and I/O schedule) absorb rows and key values not only from 6 but any number of runs, e.g., all 63 runs in this example.

With wide merging, i.e., the final merge step immediately after run generation and one full merge level, the total size of all temporary runs is $750,000 + 750,000 = 1,500,000$ rows and thus perfectly competitive with hash aggregation. As in the cost calculation for hash aggregation, the size of the original input determines the cost of each partitioning or merge level yet the size of the final output (together with memory size and partitioning fan-out or merge fan-in) determines the count of required partitioning or merge levels.

Wide merging with duplicate removal, grouping, and aggregation using an in-memory index proceeds in $O/M$ steps. Each step produces a memory load of candidate output rows, with gradual progression from one step to the next. The runs being merged must have more than $O/M$ data pages such that the key range of each data page is no larger than the key range of the in-memory index. After run generation, in each row's first temporary run on storage, the size of runs is $M$ or $2M$. In each merge level, the size increases by the fan-in $F$. The count of merge levels L must ensure that $F^L \geq O/M$ or $L \geq \log_F (O/M)$, in a striking similarity to the expression for partitioning levels required in hash aggregation.

\begin{figure}
\centering
\includegraphics{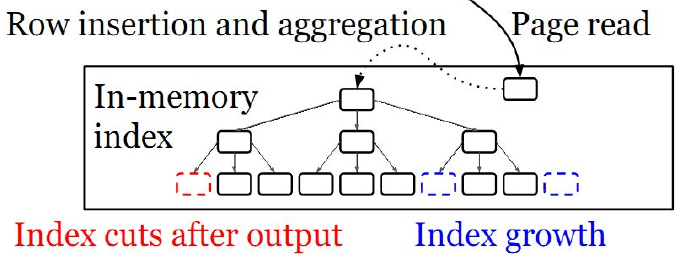}
\caption{Index expansion and contraction during wide merging.}
\label{figure_10}
\end{figure}

Figure~\ref{figure_10} illustrates this gradual progress through the key domain as required for wide merging as shown in Figure~\ref{figure_9}. A priority queue guides the page reads. For key values earlier than the top value of the priority queue, aggregates such as counts and sums are final. These can be produced as final output and removed from the in-memory index. As additional pages are read and their key values are inserted, the in-memory index grows, mostly on the right edge but occasionally also elsewhere.

\subsection{Example 4}
\label{sec_wide_merging_ex4}

Perhaps a similar example might clarify further. Compared to Example 3, this example uses more realistic values for memory size, merge fan-in, and partitioning fan-out: duplicate removal with input size (per thread) $I = 100,000,000$ rows, memory size (per thread) $M = 100,000$ rows, partitioning fan-out and (traditional) merge fan-in $F = 100$, and final output size (per thread) $O = 8,000,000$ rows. Despite more realistic values for $M$ and $F$ (compared to Example 3), it remains true that $M < O < I$.

Hash aggregation starts with a full partitioning level, which produces 100 partitions of about 1,000,000 rows each. Each partition contributes about 80,000 rows to the final output, which can be aggregated entirely in memory. The total spill volume (partitions on temporary storage) is equal to the input size or 100,000,000 rows. The algorithm requires practically the entire available memory throughout, i.e., during both partitioning and in-memory aggregation. Actually, 90\% of memory would suffice: memory size $M = 90,000$ rows and partitioning fan-out $F = 90$ suffice if the hash value distribution is perfectly uniform.

In-sort duplicate removal, grouping, and aggregation starts with run generation, which produces about 500 runs of about 200,000 rows each. Traditional merging requires 5 intermediate merge steps to reduce the count of runs from 500 to 100 as required for the final merge step\footnote{Of these, the first merge step uses fan-in 5 and the other ones use fan-in 100. These merge steps reduce the run count from 500 to 496, 397, 298, 199, and finally 100 runs.}. The output size of these steps is limited to the final output size $O = 8,000,000$ rows. Thus, the total spill volume is $I + 1 \times 1,000,000 + 4 \times 8,000,000 = 133,000,000$ rows, which is 33\% worse than the total spill volume in hash aggregation. Nonetheless, the algorithm requires the entire available memory throughout, i.e., during run generation, intermediate merge steps, and the final merge step.

In contrast, wide merging can merge and aggregate the initial 500 runs in a single step (even if memory and page sizes limit traditional merging to fan-in 100). Thus, the spill volume is equal to the input size or 100,000,000 rows, matching the performance of hash aggregation. In this algorithm, run generation requires the entire available memory (M = 100,000 rows) but the final (wide) merge step and its in-memory index require only about 40,000 rows (40\% of $M$).

Alternatively, a memory allocation for run generation of $M = 64,000$ rows reduces the size of initial runs to about 128,000 rows and produces about 782 runs. Wide merging can combine all these runs in a single final merge step using a memory allocation of about 63,000 rows. In other words, wide merging permits sort-based duplicate removal, grouping, and aggregation with equal I/O volume but with a lower memory allocation than hash aggregation.

\subsection{When to apply wide merging}
\label{sec_wide_merging_decision}

In most practical situations, hash-aggregation can process its input in memory either immediately or after one level of partitioning. In the same cases and situations, wide merging can process its initial runs, no matter their count, without the in-memory index exceeding its size limit. In other words, the situation of Section~\ref{sec_wide_merging_ex4} is much more frequent than that of Section~\ref{sec_wide_merging_ex3}. If, however, the in-memory index with candidate output rows occupies fraction $p$ of its maximal permissible size, this index and its existing rows absorb the fraction $q$ of rows in pages read from runs, and $ q < p $, then wide merging cannot absorb all rows within its memory limit.

In those cases, traditional merge steps are required before wide merging, because wide merging only applies to the final merge step. Those merge steps use their page buffers in the traditional way (one per merge input) and traditional limits on the fan-in apply. Instead of an in-memory index, the recommended in-memory data structure for these merge steps is a tree-of-losers priority queue.

Pang et~al.~\cite{PangCareyLivny-93-Sorting} investigated and experimented with techniques that let an external merge sort react gracefully to growing and shrinking memory grants. For shrinking during the merge phase, their principal technique interrupts the current merge step and introduces a preliminary merge step that reduces the count of remaining runs and thus permits resuming the interrupted merge step with a lower fan-in and thus with fewer buffers. This technique also applies to wide merging: if wide merging is not yet applicable, either from the start of a merge step or merely for a key range, a preliminary merge step can reduce the run count.

In contrast to the situations Pang et~al. investigated, wide merging in the final merge step does not require reducing the run count to the buffer count during the final merge step. Instead, it is sufficient to enable wide merging.
A preliminary merge should focus on the smallest runs, which is also the cheapest way to reduce the run count. In the context of wide merging, the significant effect of this policy is to increase the smallest size among the remaining runs, to increase the count of pages in the smallest remaining run, and thus to decrease the key range of those pages.
Those smaller key ranges per page enable wide merging, i.e., an in-memory b-tree smaller than the permissible memory might suffice for immediate duplicate removal or aggregation.
Preliminary merge steps are required until this becomes possible. As mentioned above, however, this problem should be rare, in fact precisely as rare as hash-aggregation with recursive or multi-level partitioning.

\subsection{Analysis}
\label{sec_wide_merging_analysis}

A few questions suggest themselves for further analysis. 
\begin{enumerate}
\item Is wide merging or an in-memory index useful in external merge sort without duplicate removal, grouping, and aggregation?
\item How many traditional merge levels are required before wide merging applies and becomes effective?
\item Should those earlier merge levels use priority queues (like traditional external merge sort) or should they use an in-memory index (like wide merging)?
\item What is the relationship between traditional early aggregation~\cite{BD:83} and wide merging?
\item For quickest application of wide merging, what policy should guide early merging in external merge sort for duplicate removal, grouping, and aggregation?
\end{enumerate}

In response, it seems that run generation and final merge step using an in-memory index offer performance advantages only in queries that require duplicate removal, grouping, or aggregation. Ordinary sorting, e.g., for “order by” queries and for merge join operations, does just as well with traditional algorithms and data structures, e.g., quicksort or tree-of-losers priority queues.

Wide merging is useful only in the final merge step; it might require earlier merge levels like traditional external merge sort. The number of traditional merge levels is a function of initial run size (memory size), merge fan-in, and final output size. All merge steps in which the step’s total input size is smaller than the operation’s final output size should use traditional merge logic. This analysis assumes that a merge step’s individual inputs are of similar size.

Wide merging applies when merging with a traditional fan-in processes effectively all distinct key values, i.e., when the total merge input is larger than the operation’s final output. This is precisely the first merge step (or merge level) in which traditional early aggregation~\cite{BD:83} first becomes effective. The difference is that wide merging immediately produces the operation’s final output by consuming all remaining runs, whereas traditional early aggregation still might require multiple merge steps and levels.

If traditional merge steps are required prior to wide merging in sort-based duplicate removal, grouping, and aggregation, these merge steps must create runs at least as large as the operation’s final output divided by the fan-in of traditional merge steps ($O/F$). It appears that there is little benefit in creating larger intermediate runs. Runs of size $O/F$ enable traditional early aggregation~\cite{BD:83} and, better yet, wide merging. In other words, wide merging replaces (rather than augments) traditional early aggregation. Creating runs of size $O/F$ requires $\log_F (O/M)-1$ merge levels after run generation creates runs of memory size M. With the final merge step (merge level) using wide merging, sort-based duplicate removal, grouping, and aggregation requires $\log_F (O/M)$ merge levels.

\subsection{Combining early aggregation and wide merging}
\label{sec_wide_merging_combine}

If the final output is only somewhat larger than the available memory, e.g., $O~=~2M$ or $O~=~3M$, early aggregation during run generation and its in-memory index can absorb some of the input rows without growing the index or spilling rows from memory to runs on temporary storage. For example, if $O~=~2M$, the rows in memory can absorb half of all input rows; if $O~=~3M$, the in-memory index matches and absorbs a third of all input rows; etc. Nonetheless, a large input can force many runs on temporary storage. In those cases, wide merging can eliminate one or even two merge levels. In other words, a single sort can benefit from both early aggregation and wide merging. With those two techniques and their combined effects, sort-based duplicate removal, grouping, and aggregation always performs very similarly to hash-based alternatives, as discussed further in Section~\ref{sec_perf}.

\subsection{Example 5}
\label{sec_wide_merging_ex5}

For another example that differs from Example 4 only in the final output size, consider duplicate removal with input size (per thread) I = 100,000,000 rows, memory size (per thread) M = 100,000 rows, (traditional) merge fan-in and partitioning fan-out F = 100, and final output size (per thread) O = 150,000 rows. In other words, $I \gg O = 1.5 M$.

In hash aggregation with hash-partitioning, about half of all input rows find a match in memory: hybrid hashing~\cite {Shapiro-86-Hybrid-hash} is quite effective in this case. However, the total spill volume is about $\frac{1}{2}I = 50,000,000$ rows. Sort-based aggregation with early aggregation matches the same fraction of input rows during creation of the initial runs. With replacement selection and a run size of about $2M = 200,000$ rows, about 250 initial runs are required. With run generation in read-sort-write cycles and a run size of $M = 100,000$ rows, about 500 initial runs are required. This is too much for traditional merging with fan-in $F = 100$, but nonetheless wide merging can finish the aggregation in a single merge step. Thus, this example benefits from both early aggregation and wide merging; with these techniques, sort-based aggregation can match the spill volume and performance of hash aggregation.

\subsection{Summary of wide merging}

To summarize, wide merging uses its in-memory index and a single input buffer for all runs on temporary storage. It enables the final merge step in duplicate removal, grouping, and aggregation to consume and to combine many more runs than a traditional merge step using an input buffer for each run. Wide merging applies when traditional merging would produce runs larger than the final output of the grouping operation. Matching the performance and I/O volume of hash aggregation in all cases requires both early aggregation and wide merging.

\section{Analytical comparisons of algorithms for duplicate removal} \label{sec_analysis}

Early aggregation and wide merging substantially change the competition between sort- and hash-based duplicate removal, grouping, and aggregation. These changes are pronounced both in CPU effort, often dominated by accesses to columns in rows or fields in records, and in I/O effort, which can be measured in the total size of all overflow from memory to temporary storage. The present section analyzes these two metrics, column accesses and spilling, with some surprising results. For example, sort-based duplicate removal often accesses far fewer column values than a hash-based algorithm. The subsequent section reports experimental performance measurements and comparisons.

\subsection{Column accesses} \label{sec_column_accesses}

Common wisdom holds that operations on hash values are faster than equivalent operations on complex sort keys. For example, comparing two hash values is much faster than comparing two rows with multi-column keys, in particular if some of these columns are strings or even international strings with specific locales and their sort order.

\begin{table}[h]
\centering
\begin{tabular}{|l|c|c|} 
 \hline
 \multirow{2}{*}{\textbf{Algorithmic approach}} & \multicolumn{2}{|c|}{\textbf{Output size [rows]}} \\ \cline{2-3}
 & \textbf{1} & \textbf{N} \\
 \hline\hline
  Hash-based & $3 N \times K $ & $N \times K$ \\ 
  \hline
  Sort-based & $2 N \times K $ & $2N$ to $2N \times K$ \\
  \hline
  Saving by sorting [factor] & $1.5$ & $0.5$ to $K/2$ \\
  \hline\hline
  With further optimization: & &\\
  \hline
  Sort-based w/ cache & $ N \times K$ & $N$ to $N \times K$ \\
  \hline
  Saving by sorting [factor] & $3$ & $1$ to $K$ \\
 \hline
\end{tabular}
\caption{Counts of column accesses in extreme cases.}
\label{table:column-access}
\end{table}

Offset-value coding~\cite{C:77, I:05, G:06} applied to columns can put this common wisdom into question, however. On one hand, offset-value codes are integer values compiled into the system’s source code just like hash values, meaning that a comparison of two offset-value codes is precisely as fast as a comparison of two hash values. On the other hand, sort-based query execution algorithms sometimes require far fewer column accesses than equivalent hash-based algorithms. For example, in duplicate removal for a table with $ K = 20 $ or even $ K = 200 $ columns, computing the hash values for a table of many rows, e.g., $ N = 10^6 $ rows, must access all $ N \times K $ column values. In contrast, sort-based duplicate removal and its many comparisons access only the leading columns required to distinguish each row from its neighbors in the final output. In an extreme case, if the first column happens to be unique, sort-based duplicate removal never accesses the remaining $ K - 1 $ columns. Thus, in this extreme case, sort-based duplicate removal accesses $ K $ times fewer column values than any hash-based equivalent. In another extreme case, if all rows differ from their neighbors only in the last ($ K^{th} $) key column, the count of column accesses in sort-based duplicate removal is twice that of hash-based duplicate removal. In other words, with offset-value coding, the count of column accesses in a sort-based algorithm is at worst twice that of a hash-based algorithm but possibly $ K $ times lower.

Another extreme case has no distinct rows, e.g., the output of duplicate removal is a single row. In that case, a hash-based algorithm computes a hash value for each row and then compares all columns with the one and only row in the hash table. This requires $ 3 K $ column accesses for each input row after the first one. An equivalent sort-based algorithm performs $ N - 1 $ row comparisons, each with $ 2 K $ column accesses, or less than a hash-based algorithm by a factor of $ 1.5 $.

Table~\ref{table:column-access} summarizes these extreme cases. Common cases fall in between these extreme examples. Note that $ N - 1 $ was simplified to (approximated as) $ N $ in Table~\ref{table:column-access}. Note also that the analysis of column accesses in Table~\ref{table:column-access} does not account for hash collisions, i.e., different rows mapping to the same hash value. More importantly, the sort order can be chosen freely; otherwise, hash-based algorithms would not apply. If columns with many distinct values are assigned to early positions in the sort key, hash-based duplicate removal requires more column value comparisons and column value accesses than sort-based duplicate removal. This difference is close to a factor of $ K / 2 $, which is a substantial difference for $ K = 20 $ or even $ K = 200 $ columns. Previous analyses of sort- and hash-based query execution algorithms have not considered the column count as a significant variable although Table~\ref{table:column-access} shows that it is.

The analysis of sort-based duplicate removal in Table~\ref{table:column-access} permits another improvement by a factor two, summarized at the bottom of Table~\ref{table:column-access}. If each row has a cache of all offset-value codes ever computed for that row, i.e., if no offset-value codes are ever computed redundantly, then only $ N \times K $ offset-value codes can ever be computed and the maximum number of column accesses in sort-based duplicate removal is $ N \times K $, i.e., equal to the minimum number of column accesses in hash-based duplicate removal due to the calculation of hash values.

If, due to offset-value coding, most comparisons in a sort-based algorithm are as fast as comparing hash values, the efforts to locate, access, extract, decompress, and decode $K=20$ or even $K=200$ column values in each row dwarf the effort to perform $ log_{2}(1,000,000) = 20$ comparisons for each row. In other words, the $log_{2}N$ component in the formula of comparison-based sorting matters little, because the count of column accesses dominates the overall performance of duplicate removal. Of course, as in any comparison-based sort, there are $O(NlogN)$ row comparisons, but many or most of the row comparisons are decided by offset-value codes alone, with comparisons as simple and fast as comparisons of hash values. In fact, with offset-value coding applied to columns, the number of column value comparisons is bounded to $N \times K$, i.e., it is linear in both the row count and the column count. (The alternative applies offset-value coding to bytes within a normalized key~\cite{I:05}, i.e., the entire key encoded in a single order-preserving binary string. Computing such a normalized key is just as expensive as computing a hash value from all key columns, although bytes in a normalized key could conceivably be computed on demand even if that is not possible for hash values.)

\subsection{Spilling in a memory hierarchy}

In addition to comparisons of rows, keys, and column values, the other significant cost in duplicate removal, grouping, and aggregation is moving intermediate rows up and down within the memory hierarchy. The traditional case in point is spilling from memory to temporary storage – terms commonly used in this context are external devices and external algorithms, e.g., external sorting. In many modern query execution engines and their deployments, very similar considerations and optimizations apply to spilling to memory from CPU caches, usually the second- or last-level cache.

As in all external algorithms, the pertinent questions include when to spill, what to spill, how much to spill at a time, and how much spillage is required altogether. Of course, a particularly interesting question is about the conditions for no spillage at all, i.e., when an external algorithm is not required because an internal algorithm suffices. The principal parameters in the relevant formulas are the input size $I$, the output size $O$, the memory size $M$, and the partitioning fan-out or merge fan-in $F$ (when using the entire memory) or $f$ (when using partial memory, e.g., in hybrid algorithms such as hybrid hash join~\cite {Shapiro-86-Hybrid-hash}).

The analysis below pursues these questions in detail. It omits spillage (partitioning and merging) in deep (multi-level) memory hierarchies, e.g., CPU cache – memory – external storage devices, memory – flash storage – disk storage, or CPU cache – memory – local storage – disaggregated (distributed, remote) storage. The analysis below also omits parallel algorithms including opportunistic pre-aggregation before a shuffle step (opportunistic = best effort, in-memory only; shuffle = exchange, partitioning). These omitted topics require merely straightforward extensions or repeated application of the analysis below.

The analysis below proceeds in three steps. The first step works out conditions for internal algorithms with no spillage at all. The second step assumes incredibly large inputs and outputs that require external algorithms with recursive partitioning or multi-level merging. This step is closest to a traditional “big O” complexity analysis but adds approximate constants. The third step focuses on conditions and cost functions for hybrid algorithms that divide memory between an internal algorithm for some of the input and an external algorithm for the remainder of the input. Like this last sentence, the analysis focuses on spilling from memory to external devices.

\subsubsection{No Spilling}

When a hash table is used in duplicate removal, grouping, and aggregation, only the output size governs the choice between internal and external algorithm: if the operation’s final output fits in memory, there is no need for spilling and an external algorithm. More formally, internal hash aggregation requires that $O \leq M$, independent of input size. For example, TPC-H Query~1 has four output rows for any scale factor of the database; thus, the grouping operation requires a hash table with four entries and the query is more a test of scan bandwidth than a test of the performance of duplicate removal, grouping, and aggregation.

\begin{table}[h]
\centering
\begin{tabular}{|l|c|} 
 \hline
 \textbf{Method} & \textbf{Condition for in-memory operation} \\ 
 \hline
 Hash aggregation & $O \leq M$\\
 \hline
 Sort with separate aggregation & $I \leq M$ \\ 
 \hline 
 Sort with traditional early aggregation~\cite{BD:83} & $I \leq M$ \\ 
 \hline
 New in-sort aggregation & $O \leq M$ \\
 \hline
\end{tabular}
\caption{Conditions for in-memory operations.}
\label{table:spilling-no-spilling}
\end{table}

Traditional sort-based grouping first sorts all input rows and then accumulates counts, sums, etc. for output rows. Thus, internal sorting suffices if the entire input fits in memory, or if $I \leq M$. Even with traditional early aggregation~\cite{BD:83}, sort-based duplicate removal without spilling requires that the input fits in memory. For TPC-H Query~1, traditional early aggregation might limit the size of each run on external storage to four rows but it still writes those runs and merges them over multiple merge steps and merge levels.

In contrast, the early aggregation technique proposed in Section~\ref{sec_early_agg} accumulates counts, sums, etc. immediately as input rows arrive, much in the style of hash aggregation but using an alternative format for the in-memory index. Thus, the proposed algorithms avoid spillage in the memory hierarchy under the same conditions as hash aggregation, namely if the operation’s final output fits in memory or $O \leq M$, independent of input size.

Table~\ref{table:spilling-no-spilling} summarizes those differences. The reader should consider two cases: a size reduction factor $I/O$ of $0.999$ (hardly any duplicate rows in the input) and of $1,000$ (lots of duplicate rows, e.g., after a projection on a few columns with small domains). With few duplicate rows, all algorithms require spilling at about the same operating point. With many duplicate rows in the input, traditional sort-based duplicate removal is not competitive with hash-based duplicate removal. With the new techniques for early duplicate removal (Section~\ref{sec_early_agg}), sort-based duplicate removal spills in precisely the same cases as hash-based duplicate removal.

\subsubsection{Spilling for incredibly large inputs and outputs}

The cost calculations here assume perfectly uniform hash values. Cost is measured as the total size of all partitions (in partitioning algorithms) or of all runs (in merging algorithms). Each of these runs or partitions is written once and read once, i.e., the total I/O volume is twice the costs calculated below. Note that partitioning algorithms employ random writes and sequential reads whereas merging algorithms employ random reads and sequential writes – the cost analysis below ignores any effects of this difference.

For extremely large inputs, hash aggregation partitions its input recursively. Each partitioning level moves the entire input; each partitioning step moves data from one input to multiple outputs. The number of outputs from a single input is the partitioning fan-out $F$; with memory used for buffers such that multiple rows can be moved in efficient units of reading and writing, the fan-out is limited by the available memory size divided by the sizes of pages or units of data movement within the memory hierarchy, i.e., $F = M/P$ for page size $P$.

With uniform hash values, each partitioning level reduces partition sizes by a factor $F$. Partitioning continues until each partition can be processed in memory. It is not required that the partition fit in memory, only that the output after duplicate removal, grouping, and aggregation be smaller than memory. Thus, the number of partitioning levels for a very large input is $log_{F}(O/M)$, independent of input size. The size of all partitions over all recursive partitioning levels is thus $log_{F}(O/M) \times I$. The output size governs the count of partitioning levels; the input size governs the cost of each partitioning level.

As mentioned earlier, some implementations of hash aggregation use partitioning only in memory, e.g., a hash table accumulates counts, sums, etc. immediately as input arrives, but external merge sort is used for spilling. The sort key may or may not include the hash value; if it does, the hash value typically is an artificial leading key column. Nonetheless, the total cost of spilling is determined by sort-based algorithms and merging, as analyzed next.

In traditional external merge sort with subsequent duplicate removal, the input size determines the total cost of spilling as the product of the count of merge levels and the cost per merge level. The merge fan-in $F$ is the same as the partitioning fan-out in recursive partitioning, i.e., the available memory size divided by the size of pages or units of data movement within the memory hierarchy, i.e., $F = M/P$. 

Once the merge fan-in is set, the total size of all runs on external storage is $log_{F}(I/M) \times I$. When duplicate removal is a separate step applied only after sorting, the output size of duplicate removal has no effect on the cost of sorting.

External merge sort with traditional early duplicate removal~\cite{BD:83} requires separating initial merge levels and remaining optimized merge steps (note the distinction of merge levels versus merge steps). The transition occurs when merged runs could be larger than the final output $O$ but are limited to $O$ by traditional early duplicate removal. The size of individual runs is $M$ after run generation and $M \times F^L$ after $L$ merge levels. This size exceeds the final output size O after $L \geq log_{F}(O/M)$ initial merge levels. The total size of all runs created in these initial merge steps is $log_{F}(O/M) \times I$, because each initial merge level moves the entire input. Note the similarity of these initial merge costs to the total costs of recursive hash partitioning.

In duplicate removal, the average count of duplicate rows equals the quotient of input and output sizes, i.e., $I/O$. When $I/O$ runs remain, traditional early duplicate removal ensures that all further merge steps have output size $O$. Each merge step reduces the count of runs by $F - 1$; the goal is to reduce the number of remaining runs from $I/O$ to $1$, which requires $(I/O - 1)/(F - 1)$ merge steps with fan-in $F$ and output size $O$. The total size of all runs created in these remaining merge steps is $(I/O - 1)/(F - 1) \times O = (I - O)/(F - 1)$. The combined cost of full merge levels plus optimized merge steps is $log_{F}(O/M) \times I + (I - O)/( F - 1)$.

Wide merging as proposed in Section~\ref{sec_wide_merging} performs the same initial merge levels but condenses all remaining merging into a single step with a fan-in of $I/O$ and thus exceeding $F$. Wide merging applies when an additional input run does not increase the output size or, equivalently, when an $F$-way merge produces all key values. For example, if the memory size is $M = 1$ (i.e., memory size is the unit to measure other sizes), the merge fan-in is $F = 10$, and each input run size is $\frac{1}{8} O$, then the merge output size is equal to the operation’s final output size $O$, not $10 \times \frac{1}{8} \times O = 1\frac{1}{4} \times O$. Importantly, the output size is $O$ no matter how many additional runs the merge step consumes, i.e., the actual merge fan-in may far exceed the traditional maximal fan-in $F$. Thus, with wide merging, the total size of all runs created is $log_{F}(O/M) \times I$.

The analysis above of sort-based duplicate removal, grouping, and aggregation assumes run generation with initial runs of size equal to memory $M$, which is typical for run generation using read-sort-write cycles and quicksort. Replacement selection produces longer runs, typically twice as large as memory~\cite{G:67}, with about one additional run due to start-up and shut-down of the replacement selection logic. Thus, the expected run count is $I/(2M)+1$ and the merge cost in external merge sort is $log_{F}(1+I/(2M)) \times I$. With wide merging, the expected merge cost is $log_{F}(1+O/(2M)) \times I$. The savings are about $log_{F}(2) = 1/log_{2}(F)$ merge levels, which is a meaningful difference only for a fairly small value of the merge fan-in, e.g., $F=8$. It is negligible for typical contemporary choices for the maximal merge fan-in, e.g., $100$ or even $1,000$. Run generation by replacement selection might be interesting, however, for hybrid algorithms as analyzed below.

\begin{table}[h]
\centering
\begin{tabular}{|l|c|} 
 \hline
 \textbf{Method} & \textbf{Total spill volume [bytes]} \\ 
 \hline
 Hash aggregation with partitioning & $log_{F}(O/M) \times I $\\
 \hline
 Sort with separate aggregation & $log_{F}(I/M) \times I$  or  $log_{F}(1+I/(2M)) \times I$ \\ 
 \hline 
 Sort with traditional early aggregation & $log_{F}(O/M) \times I + (I - O)/(F - 1)$ \\ 
 \hline
 New in-sort aggregation & $log_F(O/M) \times I$ or $log_{F}(1+O/(2M)) \times I$ \\
 \hline
\end{tabular}
\caption{Total spill volumes.}
\label{table:spilling-large-output}
\end{table}

Table~\ref{table:spilling-large-output} summarizes the spilling costs for duplicate removal with input and output incredibly large. The simplest formula applies to sorting with separate, subsequent duplicate removal: it is the traditional $N log N$ formula well known for divide-and-conquer algorithms such as multi-level external merge sort. The cost formulas for sorting with separate duplicate removal assume run generation with either read-sort-write cycles (e.g., using quicksort) or replacement selection (e.g., using a priority queue); the latter produces almost half as many initial runs on temporary storage. The cost formula for hash aggregation with partitioning is the same as external merge sort except that the output size, not the input size, determines the count of partitioning levels. Duplicate removal by sorting with traditional early aggregation spills as much as hash aggregation with partitioning before the optimized merge step, which adds no more than a fraction of the original input size. Wide merging (Section~\ref{sec_wide_merging}) eliminates even that – thus, the cost for spilling and moving data up and down within the memory hierarchy is the same as in hash aggregation with partitioning. The remaining difference in database query processing is, of course, interesting orderings with its advantages in complex query execution plans. With run generation by replacement selection, duplicate removal by sorting with wide merging gains a slight edge over all other methods, including hash aggregation.

\subsubsection{Hybrid algorithms}

Hybrid algorithms, e.g., hybrid hash join and hybrid hash aggregation~\cite {Shapiro-86-Hybrid-hash}, divide memory between an internal algorithm for some of the input and an external algorithm for the remainder of the input. For example, if half the build input of a hybrid hash join fits in memory, half of both inputs does not require any spilling. Similarly, if half the output of a hybrid hash aggregation fits in memory, half of the input does not require any spilling. In sort-based duplicate removal, half of memory might use an in-memory index for immediate duplicate removal while the other half of memory is used for run generation using quicksort, a priority queue, or an in-memory index.

\begin{table}[h]
\centering
\begin{tabular}{|l|c|c|} 
 \hline
 \textbf{Method} & \textbf{Actual fan-out or fan-in} & \textbf{Condition for hybrid operation} \\ 
 \hline
 Hash aggregation with partitioning & $f = \ceil*{(O - M)/(M - P)}$ & $M < O \leq F \times M = M^2/P$ \\
 \hline
 Sort with separate aggregation & $f = \ceil*{(I - M)/(M - P)}$ & $M < I \leq F \times M = M^2/P$ \\ 
 \hline 
 New in-sort aggregation & $f = \ceil*{(O - M)/(M - P)}$ & $M < O \leq F \times M = M^2/P$ \\
 \hline
\end{tabular}
\caption{Condition for hybrid operation}
\label{table:spilling-hybrid}
\end{table}

In all cases, the hybrid approach is promising only if the spilled data can be processed efficiently, i.e., without recursive partitioning or multi-level merging. Thus, hybrid algorithms are limited to input size less than “the square of memory”~\cite{Shapiro-86-Hybrid-hash}. More precisely, hybrid hash join applies if the build input size is less than $M^2/P = F \times M$ for page size $P$ and partitioning fan-out $F$. Hybrid hash aggregation applies if the operation’s final output size is less than “$M^2$” or, more precisely, $O \leq F \times M$. Table~\ref{table:spilling-hybrid} summarizes the “hybrid” considerations for hash aggregation with partitioning, sort with separate aggregation, and sort with early aggregation and wide merging. 

More specifically, if hash-based duplicate removal employs a fan-out of $f$, with $1 \leq f \leq F$, the fraction $f/F$ of memory is used as output buffer for f partitions on external storage. The goal is to create partitions that each produce output volume $M$ in a single step such that this output can be accumulated using an in-memory hash table. Each output partition of size $M$ corresponds to an input volume of $M \times I/O$. The total spill volume $f \times M \times I/O$; the total final output volume from all spilled partitions is $f \times M$. The remaining memory holds a hash table. This hash table produces $(1 - f/F) \times M$ of the output $O$ after absorbing input volume $(1 - f/F) \times M \times I/O $. Thus, $f$ should be the smallest integer such that $O \leq f \times M + (1 - f/F) \times M$ or $O/M - 1 \leq f - f/F = f(1 - 1/F)$ or $f \geq (O/M - 1)/(1 - 1/F) = (O - M)/(M - P)$. Alternatively, $f$ must be chosen such that $I \leq f \times M \times I/O +(1 - f/F) \times M \times I/O$ or $I/(M \times I/O) \leq f+(1 - f/F)$ or $O/M - 1 \leq f - f/F = f(1 - 1/F)$ or $f \geq (O/M - 1)/(1 - 1/F) = (O - M)/(M - P)$. Pragmatically, the actual partitioning fan-out should be set to $f=\ceil*{(O - M)/(M - P)}$. This value falls into the permissible range $f \leq F$ if the output size satisfies $O \leq F \times M$.

For a specific example, assume $I=275 M$, $F=10$, $O=2\frac{3}{4} M$, and thus $I/O=100$ and $f=2$. In other words, the fraction $f/F=2/10$ of memory is used as partitioning output buffer and the remaining $8/10$ of memory for an in-memory hash table. With final output size $2\frac{3}{4}=11/4$ of the memory size and the hash table $4/5$ of memory size, the hash table can accumulate $(4/5)/(11/4) \approx 3/11$ of the output whereas the external partitions temporarily store $\sim8/11$ of the input or $\sim200 \times M$.

The extreme examples of hybrid hash aggregation, i.e., $f=0$ or $f=F$, may be considered not hybrid at all. If $O \leq M$, in-memory duplicate removal suffices, which could be seen as “hybrid” hash aggregation with the actual fan-out $f=\ceil*{(O - M)/(M - P)}=0$. If $(F - 1) \times M < O \leq F \times M$, $f=\ceil*{(O - M)/(M - P)}=F$ is not a true hybrid but full partitioning with maximal fan-out $F$ and no immediate in-memory duplicate removal.

For duplicate removal, grouping, and aggregation based on or built into external merge sort, hybrid run generation divides memory into an in-memory index to detect and remove duplicate rows immediately and a workspace to prepare sorted runs on temporary storage for a single, final merge step. Actually, a single in-memory index can serve both purposes. This choice requires assigning some key values (or ranges) to remain in memory and the remainder to spill to runs on temporary storage.

Hybrid run generation may use the same division of memory as found optimal in hash aggregation: the memory fraction $f/F$ is used for run generation, whereas the remaining memory fraction $1 - f/F$ supports an index for immediate in-memory duplicate removal. This in-memory index produces $(1 - f/F) \times M$ of the output, i.e., it absorbs input volume $(1 - f/F) \times M \times I/O$. These fractions equal those determined above for hybrid hash aggregation. With run generation using fraction $f/F$ of memory, each run is of size $f/F \times M$.

Wide merging can merge more than $F$ runs in a single merge step. If $F+1$ runs produce no more output (after duplicate removal) than $F$ runs, wide merging applies and can merge any number of runs. This consideration applies to any key range; thus, the fact that hybrid run generation has absorbed some key range during the input phase is irrelevant. In other words, wide merging as a single, final merge step applies if $F$ runs of size $M$ produce the entire output of size $O$, i.e., $O \leq F \times M$. This is precisely the same condition required for hybrid hash aggregation.

Using the prior example with $F=10$, $O=2\frac{3}{4} \times M$, $I=100 \times O$, and $f=2$, hybrid run generation uses $8/10$ of memory for an in-memory index and $2/10$ of memory for run generation. The in-memory index accumulates $3/11$ of the output whereas the external runs temporarily store $8/11$ of the input. With $I=275 \times M$, total spilling is $200 \times M$, and with an expected average run size $f/F \times M=0.2M$, there will be $1,000$ runs. Traditional merging with fan-in $F=10$ requires three merge levels and $111$ merge steps. With $O \leq F \times M$, however, a single wide merge step suffices. Thus, the total data volume moved up and down within the memory hierarchy is the same for hybrid hash aggregation and for hybrid run generation followed by wide merging.

With run generation by replacement selection (rather than read-sort-write cycles), the expected run size is twice the memory dedicated to run generation. The combination of replacement selection and wide merging permits hybrid run generation and a single wide merge step in additional cases, namely whenever $O \leq F \times 2M$. Thus, there are cases in which this combination can process its entire input by spilling each input row once whereas hash aggregation needs to spill some of its input twice.

\subsection{Summary of analytical comparisons}

To summarize, this analysis and comparison of sort- and hash-based algorithms for duplicate removal, grouping, and aggregation considers both CPU effort and I/O effort.
With respect to CPU effort, sort-based algorithms require more row comparisons than equivalent hash-based algorithms but often fewer column comparisons. Moreover, with offset-value coding, they often require far fewer column accesses. In fact, if all offset-value codes are cached once computed, the worst-case count of column accesses in sort-based duplicate removal equals the best-case count of column accesses in hash-based duplicate removal, and the best-case count of column accesses in sort-based duplicate removal is substantially better. With respect to I/O effort, sort-based algorithms, with the new techniques for early aggregation and wide merging, require no more spilling to temporary external storage than equivalent hash-based algorithms.
Thus, this algorithm analysis justifies our belief that a 

\section{Experimental comparisons}
\label{sec_perf}single algorithm is sufficient for all cases of grouping rows from unsorted inputs.

\subsection{Product Implementation} \label{sec_impl}

This section briefly summarizes the implementation of the new in-sort aggregation techniques for Google's F1~Query~\cite{S+:18, S+:13}. F1~Query is a federated query processing platform that executes SQL queries against data stored in different storage systems at Google, e.g., BigTable~\cite{C+:08}, Spanner~\cite{C+:13}, Mesa~\cite{G+:16}, and others. Before this work, F1~Query had two aggregation operators. First, for sorted input, in-stream aggregation requires little CPU effort and hardly any memory. The F1~Query optimizer chooses in-stream aggregation whenever possible based on interesting orders in the input~\cite{SAC:79}. Second, for unsorted input, hash-based aggregation relies on an in-memory hash table. This hash-based operator relies on external merge sort when the output is larger than the available memory allocation.

The new in-sort aggregation algorithm reuses the row-plus-row accumulation component of hash-based and in-stream aggregation. A new order-based indexing component is used for detecting duplicates and groups. For each input batch, the operator first sorts the batch, usually within the CPU cache as these batches are small, to detect duplicates within a batch. Only distinct key values within a batch are looked up in the ordered index with the guided search technique. When running out of memory, the ordered index guides the sequence of rows spilled to intermediate storage, creating sorted runs. These runs are eventually merged and aggregated using wide merging. Contrary to our initial design, the current implementation uses read-sort-write cycles, not replacement selection; therefore, the size of initial runs equals the memory size, even if each run might have absorbed substantially more input with replacement selection in duplicate removal and grouping.

As the new in-sort aggregation produces sorted output as a byproduct of using the ordered index, an optimizer can take advantage of this property. For example, in queries with a “group by” clause followed by an “order by” clause with the sort keys matching the grouping keys, the F1~Query optimizer avoids redundant sorting. Before our new operator, F1~Query planned such aggregation queries using either a hash aggregation followed by a sort or a sort followed by an in-stream aggregation. Plan choices can be sub-optimal due to missing or inaccurate cardinality information. The new in-sort aggregation operator overcomes this problem by always enabling the optimal plan.

The present section reports on the performance of the new sort-based grouping algorithm in F1~Query in order to support or refute our expectations and hypotheses about the new in-sort grouping techniques. With no innovation in parallel query execution, all experiments here report local or single-threaded efficiency, scalability, and robustness or reliability of performance. There are four groups of performance results. The first group of experiments replicates earlier examples. The second group focuses on early aggregation using an in-memory index for run generation. The third group focuses on wide merging using an in-memory index during the final merge step. The fourth group of performance results replicate and augment an earlier comparison of sort- and hash-based duplicate removal, grouping, and aggregation. 
All algorithms are implemented and tuned for production. 
All experiments below ran on a workstation with a fast local storage device; details are omitted on purpose.

\subsection{Validation of examples}

\begin{figure}
\centering
\begin{minipage}{.5\textwidth}
  \centering
  \includegraphics{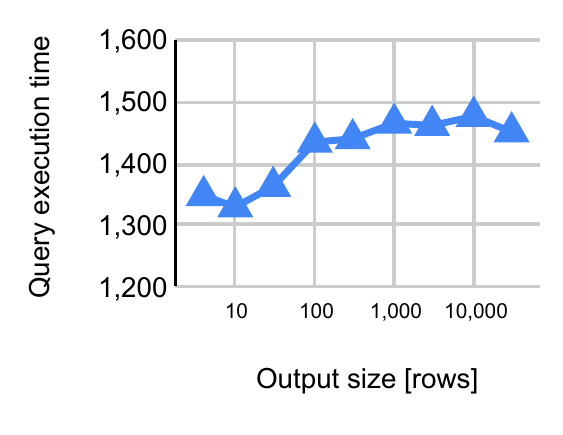}
  \captionof{figure}{In-memory grouping using a b-tree index.}
  \label{figure_11}
\end{minipage}%
\begin{minipage}{.5\textwidth}
  \centering
  \includegraphics{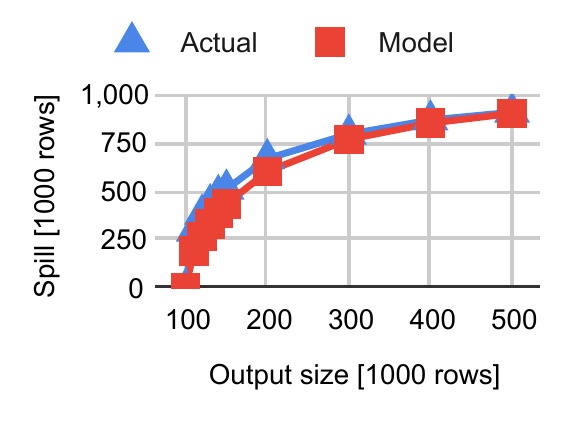}
  \captionof{figure}{Spill volume to runs on temp. storage.}
  \label{figure_12}
\end{minipage}
\end{figure}

Example 1 (Section~\ref{sec_early_agg_ex1}) focuses on TPC-H Q1, i.e., a grouping query with a final output smaller than the available memory allocation ($O \leq M$). Figure~\ref{figure_11} shows the performance of in-memory grouping and aggregation using an in-memory b-tree index. From left to right, the output size varies from 4 to 30,000 rows. The input size is constant 6,000,000 rows. As is readily apparent, the CPU effort is low and fairly consistent, because any effects due to the logarithmic depth of the ordered index vanishes compared to other CPU efforts in the query evaluation plan.

Example 2 (Section~\ref{sec_early_agg_ex2}) also varies TPC-H Q1 with output sizes beyond memory size ($O>M$ but $O<F \times M$). Figure~\ref{figure_12} compares the total size of initial runs to a model that assumes run generation with replacement selection and computes the spill volume as $M + (1 - M/O) \times I$. In contrast, our implementation relies on run generation by read-sort-write cycles. Given this difference, the distance between these curves seems acceptable.

\begin{figure}
\centering
\begin{minipage}{.5\textwidth}
  \centering
  \includegraphics{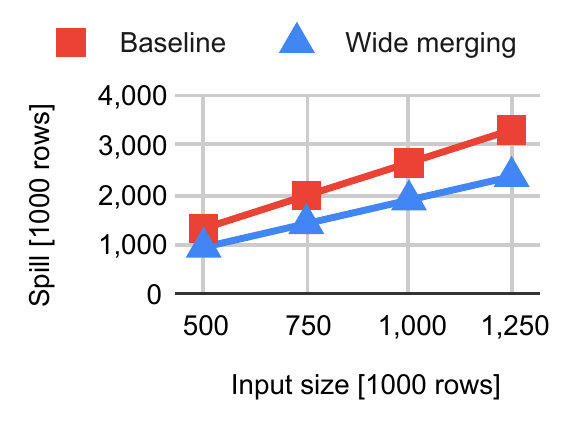}
  \captionof{figure}{Multiple merge levels.}
  \label{figure_13}
\end{minipage}%
\begin{minipage}{.5\textwidth}
  \centering
  \includegraphics{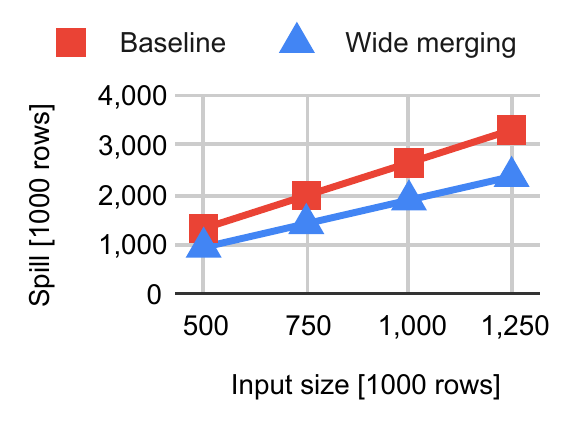}
  \captionof{figure}{Effect of wide merging.}
  \label{figure_14}
\end{minipage}
\end{figure}

Example 3 (Section~\ref{sec_wide_merging_ex3}) assumes tiny memory, merge fan-in, and partitioning fan-out. Therefore, all algorithms incur multiple partitioning or merge levels. In contrast to traditional sorting and merging, wide merging limits the merge depth $\log_F (O/M)$ versus $\log_F (I/M)$. Figure~\ref{figure_13} shows the performance advantage of wide merging over aggregation while writing runs~\cite{BD:83} as baseline: entire merge levels can be avoided, whereas earlier method merely reduce the size of intermediate runs on temporary storage.

Example 4 (Section~\ref{sec_wide_merging_ex4}) assumes realistic memory size, merge fan-in, and partitioning fan-out. Therefore, a single level of merging suffices if wide merging is available. Figure~\ref{figure_14} shows that even for modest input sizes, traditional early aggregation (the baseline) requires multiple merge levels. Note that an algorithm can spill more rows than its input size only if it spills some rows multiple times. In contrast, each input row spills only once when wide merging combines all runs in the first (and only) merge step immediately after run generation.

Example 5 (Section~\ref{sec_wide_merging_ex5}) shows that in some cases, early aggregation and wide merging are both required for best performance of sort-based duplicate removal, grouping, and aggregation. The experiment in Figure~\ref{figure_18} (Section~\ref{sec_perf_wide_merging}) confirms the example calculations.

In summary, the examples and the related experiments demonstrate that early aggregation and wide merging, by using in-memory ordered indexes instead of the traditional priority queues, derive substantial benefits for duplicate removal, grouping, and aggregation.

\subsection{Early aggregation during run generation}

The next set of experiments and diagrams focuses on the hypotheses that, for any input size, output size, row size, page size, and memory size,
\begin{enumerate}
\item an ordered in-memory index can be as efficient as a hash table;
\item an in-memory index permits run generation as efficient as quicksort and priority queues; and
\item requirements for temporary storage are the same for an ordered index and run generation as for hash table and hash partitioning.
\end{enumerate}
The experiments cannot claim to cover all sizes and key values distributions, but they may nonetheless helpful in understanding the performance and scalability of in-sort aggregation with early aggregation during run generation.

\begin{figure}
\centering
\begin{minipage}{.5\textwidth}
  \centering
  \includegraphics{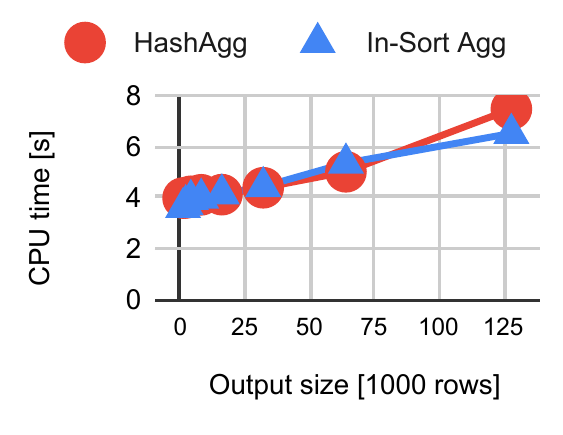}
  \captionof{figure}{Performance of in-memory indexes.}
  \label{figure_15}
\end{minipage}%
\begin{minipage}{.5\textwidth}
  \centering
  \includegraphics{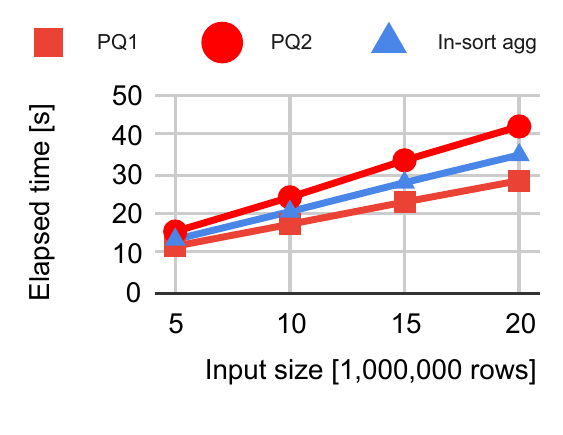}
  \captionof{figure}{Performance of run generation. ``PQ1'' and ``PQ2'' use tree-of-losers priority queues optimized with offset-value coding and normalized keys respectively.}
  \label{figure_16}
\end{minipage}
\end{figure}

Figure~\ref{figure_15} shows the performance of in-memory aggregation using either a hash table (hash aggregation) or a b-tree (in-sort aggregation). None of these experiments spill to temporary storage. As is readily apparent, the performance of hash table and in-memory b-tree, both properly optimized, is quite similar. Their remaining performance differences are so minor that other query execution costs such as aggregation arithmetic dominate them. In that sense, in-memory b-trees in sort-based grouping are as fast as hash tables in hash-based grouping, supporting Hypothesis 1.

Figure~\ref{figure_16} shows the performance of three implementations of run generation. Two of these use tree-of-losers priority queues; one of them is optimized with normalized keys for fast comparisons and poor man's normalized keys for cache efficiency~\cite{G:06}; the other priority queue is optimized with offset-value coding~\cite{C:77}. The third implementation of run generation uses an in-memory b-tree. Again, other query execution costs such as aggregation arithmetic dominate these differences and the experiment supports Hypothesis 2.

Figure~\ref{figure_17} shows the number of initial runs spilled to temporary storage. Each run is the size of memory, either a hash table sorted and written as run or a b-tree written in total when memory is full. Recall that the hash aggregation in F1~Query uses what Boncz et~al.~\cite{BNE:13} call ``hash-based early aggregation in a sort-based spilling approach,'' which sorts rows in an overflowing hash table, writes them as initial runs on temporary storage, merges those runs, and applies duplicate removal, grouping, and aggregation only during the final merge step. Recall also that our implementation of in-sort aggregation uses its in-memory index for run generation in read-sort-write cycles, not replacement selection. Thus, the counts of initial runs are practically equal, supporting Hypothesis 3.

\subsection{Wide merging in the final merge step}
\label{sec_perf_wide_merging}

The next experiments test the hypotheses that, for any input size, output size, row size, page size, and memory size,
\begin{enumerate}
\setcounter{enumi}{3}
\item wide merging combines many more runs than traditional merging and thus can avoid entire merge levels from traditional sort-based algorithms for duplicate removal, grouping, and aggregation;
\item sorting after aggregation can be as expensive as the aggregation such that in cases of equal “group by” and “order by” lists, sort-based aggregation can cost half of hash aggregation plus sorting;
\item sort-based aggregation can process a “count (A), count (distinct A)” query with grouping in a single sort using both early aggregation and wide merging, whereas hash-based query processing requires two hash aggregation operations – the performance difference can equal a factor two.
\end{enumerate}

\begin{figure}
\centering
\begin{minipage}{.5\textwidth}
  \centering
  \includegraphics{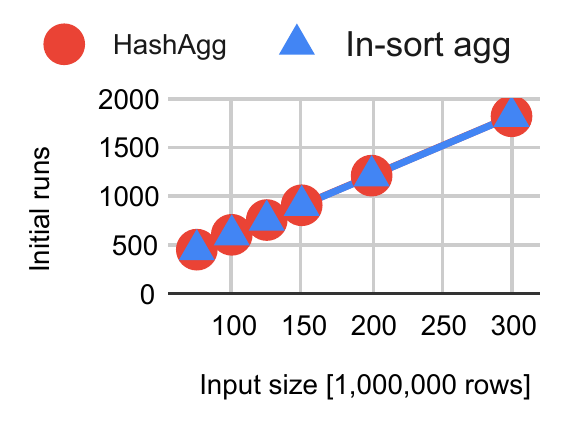}
  \captionof{figure}{Count of runs spilled from memory to temporary storage.}
  \label{figure_17}
\end{minipage}%
\begin{minipage}{.5\textwidth}
  \centering
  \includegraphics{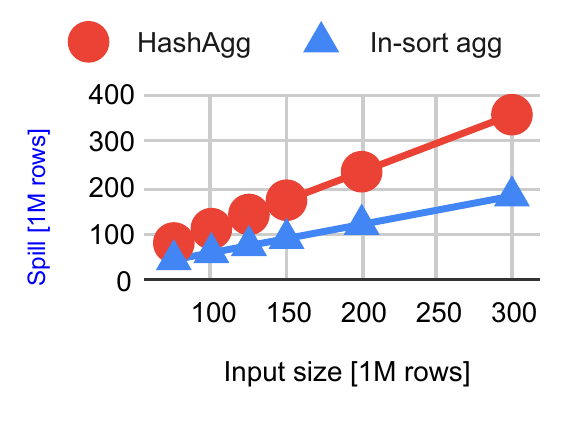}
  \captionof{figure}{Spill volume.}
  \label{figure_18}
\end{minipage}
\end{figure}

Figure~\ref{figure_18} reports on the total size of all runs for the experiment of Figure~\ref{figure_17}. Note that this experiment compares optimized in-sort aggregation with the original hash aggregation of F1~Query. Both algorithms spill from memory to sorted runs on temporary storage. The two algorithms achieve the same amount of in-memory aggregation during this phase, and thus merging in the two algorithms starts with the same counts and sizes of partially aggregated runs. The original algorithm of F1~Query relies on traditional merging, which requires multiple merge steps with intermediate merge results. Thus, the total spill volume exceeds the input size for all input sizes in Figure~\ref{figure_18}. In contrast, the new algorithm employs wide merging for duplicate removal, grouping, and aggregation. A single merge step suffices and no intermediate merge steps create any additional spill volume. For all input sizes in Figure~\ref{figure_18}, the total spill volume is much less than the input. One of the data points precisely matches Example 5 (Section~\ref{sec_wide_merging_ex5}) and all data points support Hypothesis 4.

\begin{figure}
\centering
\begin{minipage}{.5\textwidth}
  \centering
  \includegraphics{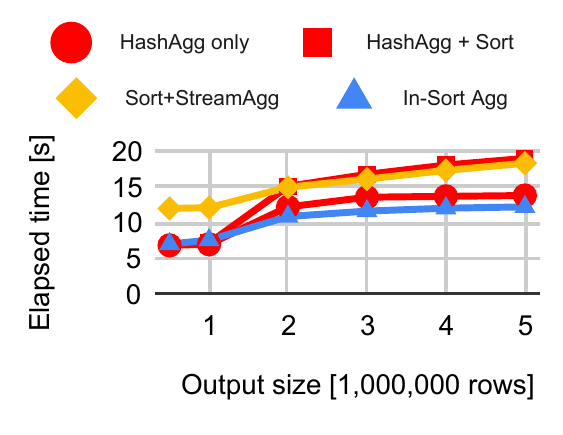}
  \captionof{figure}{Cost of sorting after aggregation.}
  \label{figure_19}
\end{minipage}%
\begin{minipage}{.5\textwidth}
  \centering
  \includegraphics{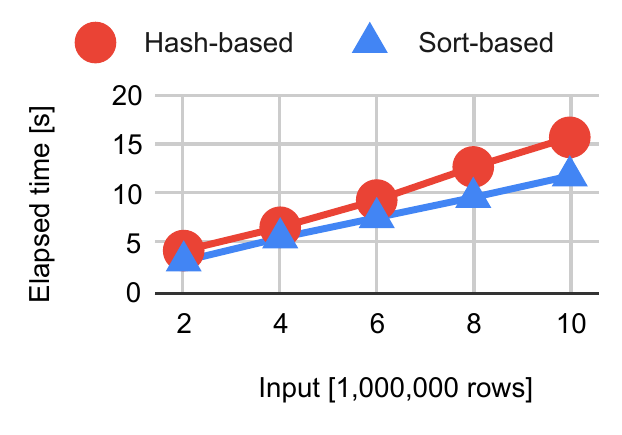}
  \captionof{figure}{Cost of “count” and “count distinct” queries.}
  \label{figure_20}
\end{minipage}
\end{figure}

Figure~\ref{figure_19} shows the cost of a query with matching “group by” and “order by” clauses over a table of $I = 6,000,000$ rows. If the output of an initial duplicate removal is small, in particular no larger than memory $M = 1,000,000$ rows, the cost of a subsequent sort operation barely matters. If, however, the intermediate result is large, then satisfying both clauses with a single operation is very beneficial, supporting Hypothesis 5.

Figure~\ref{figure_20} shows the cost of duplicate removal with subsequent grouping. With hash-based algorithms, two hash tables (and possibly overflow to temporary storage) are required. With a sort-based algorithm, a single sort can perform the duplicate removal using an interesting ordering for the subsequent grouping. Thus, only one memory-intensive operation is required with savings up to a factor of two, supporting Hypothesis 6.

\subsection{Effects of interesting orderings}

The next experiment tests the hypotheses that:
\begin{enumerate}
\setcounter{enumi}{6}
\item interesting orderings are important not only for b-tree scans and merge joins but also for query evaluation plans with duplicate removal, grouping, and aggregation; and
\item SQL set operations such as “intersect” can be much faster using sort-based query plans than using hash-based query plans. 
\end{enumerate}

\begin{figure}
\centering
\begin{minipage}{.5\textwidth}
  \centering
  \includegraphics{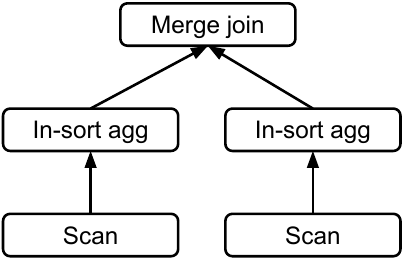}
  \captionof{figure}{Sort-based plan for “intersect distinct”.}
  \label{intersect_mj_plan}
\end{minipage}%
\begin{minipage}{.5\textwidth}
  \centering
  \includegraphics{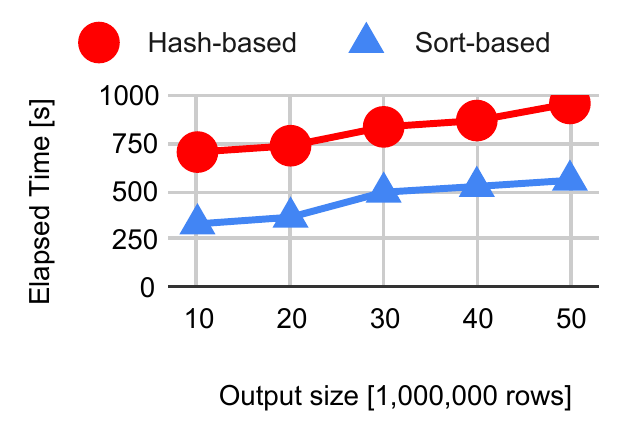}
  \captionof{figure}{Cost of “intersect distinct”.}
  \label{figure_intersect}
\end{minipage}
\end{figure}

Figure~\ref{intersect_mj_plan} shows a query evaluation plan for a very simple SQL query computing the intersection of two tables, e.g., “select B from T1 intersect select B from T2”. If column B is not a primary key in tables T1 and T2, correct execution requires duplicate removal for each input plus a join algorithm to find intersection result. Without the work described in this paper, F1~Query chooses a hash-based physical plan with three blocking operators: two hash aggregation operators for duplicate removal and the hash join for set intersection. 
In contrast, with the new in-sort aggregation operator, F1~Query chooses a sort-based plan physical plan with only two blocking operators: both are in-sort aggregation operators for duplicate removal. Since the output of in-sort aggregation operators are sorted on column B, F1~Query exploits interesting ordering and uses merge join for set intersection, supporting Hypothesis 7.

Figure~\ref{figure_intersect} shows the performance of sort- and hash-based plans for this query. Each input table has  $I =$ 100,000,000 rows; the memory for each operator is $M =$ 10,000,000 rows. In the hash-based plan, both duplicate removal operations and the join might spill to temporary storage; each input row is spilled twice. In contrast, the sort-based plan spills each input row at most once. Thus, the effort for spilling is cut in half due to interesting orderings, supporting Hypothesis 8. With in-sort aggregation, set intersection and its most efficient query evaluation plans benefit not only users' “intersect” queries but also star queries and snowflake queries in relational data warehousing.

\subsection{A belated correction}

Section 4.4 and Figure~11 of~\cite{G:93} compare sort- and hash-based duplicate removal, grouping, and aggregation. The overall conclusions are that sorting the input for subsequent in-stream aggregation is not competitive and that both sort- and hash-based aggregation exploit strong data reduction and small output sizes.

\begin{figure}
\centering
\includegraphics{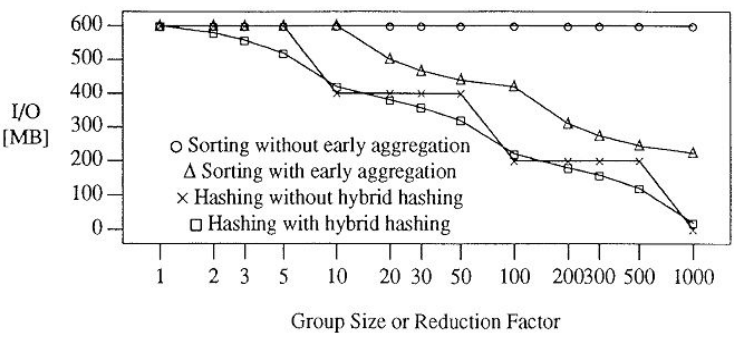}
\caption{Prior algorithm comparison~\cite{G:93}.}
\label{figure_21}
\end{figure}

Figure~\ref{figure_21} is a copy of Figure~11 of~\cite{G:93}. As perhaps appropriate at the time, the experimental parameters are input size $I = 100MB$, memory size $M = 100KB$, page size $P = 8KB$, merge fan-in and partitioning fan-out $F = 10$, and output size $O$ varying from 100MB to 100KB, or from input size $I$ to memory size $M$. The “group size or reduction factor” is the quotient of input and output sizes, $I \div O$. “Early aggregation” in this diagram means duplicate removal within runs on temporary storage~\cite{BD:83}. The I/O volume reflects both writing and reading on temporary storage, i.e., the values in Figure~\ref{figure_21} are $2 \times$ higher than the “total run size” metric used in the present paper.

\begin{figure}
\centering
\includegraphics{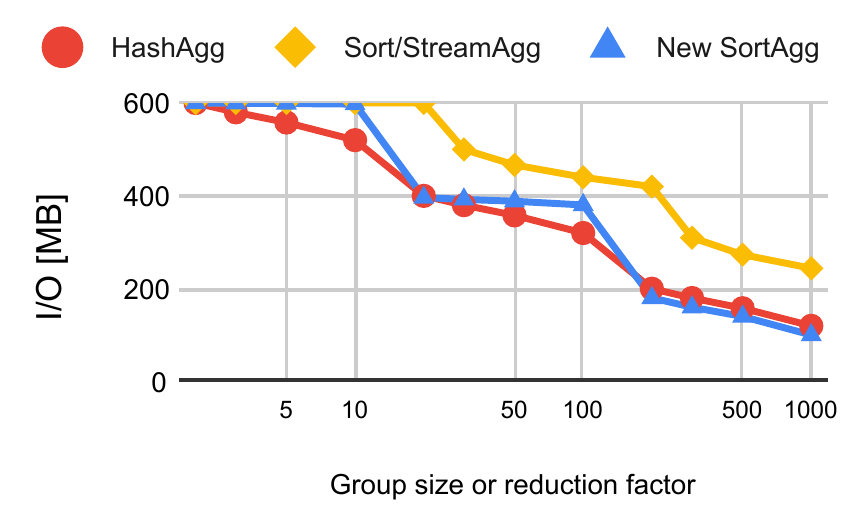}
\caption{Revised algorithm comparison.}
\label{figure_22}
\end{figure}

Figure~\ref{figure_22} augments Figure~\ref{figure_21} and maybe should replace Figure 11 of~\cite{G:93}, if that were possible. Compared to Figure~\ref{figure_21}, Figure~\ref{figure_22} omits the curves for sorting without early aggregation and for hash aggregation without hybrid hashing, but it reproduces two of the curves using the same cost functions and parameters as in earlier work~\cite{G:93}. In addition to sorting with traditional early aggregation~\cite{BD:83} and hash aggregation with hybrid hashing, Figure~\ref{figure_22} shows a new curve for sort-based aggregation with both early aggregation (run generation using an in-memory index – Section~\ref{sec_early_agg}) and wide merging (a final merge step using an in-memory index – Section~\ref{sec_wide_merging}). The essential observation is that the gap between sort- and hash-based aggregation, clearly visible in Figure~\ref{figure_21}, practically disappears in Figure~\ref{figure_22}. The curves are particularly close in the operating range towards the right with only a single merge or partitioning step. With today’s memory sizes, most grouping operations in production workloads require only one merge or partitioning step. Put differently, sort- and hash-based aggregation algorithms perform very similarly for unsorted inputs. While a single diagram cannot prove it, this is true for any combination of input and output sizes. Moreover, sort-based aggregation is less susceptible to skew in the key value distribution than hash aggregation is to skew in the hash value distribution.

\subsection{Summary of hypotheses and observations}

Together, our experiments confirm the calculations in our earlier examples, support our claims and hypotheses about the effectiveness of sort-based aggregation with early aggregation and wide merging, and belatedly correct an algorithm comparison published a quarter century ago.

\section{Summary and conclusions}

In summary, traditional sort-based algorithms for duplicate removal, grouping, and aggregation are not quite competitive with hash-based query execution. Reflecting the current common wisdom, M\"{u}ller et~al.~\cite{MSL:15} state that “hashing allows for early aggregation while sorting does not.” The techniques introduced in Sections~\ref{sec_early_agg} and~\ref{sec_wide_merging} correct this deficiency, as shown in Sections~\ref{sec_analysis} and~\ref{sec_perf}. 
Our analysis focuses on column accesses as dominant CPU cost and I/O as dominant data movement cost, reflecting our experience with industrial systems and correlating with our performance measurements.
For small outputs, early aggregation uses an in-memory index during run generation using read-sort-write cycles or replacement selection, spilling to temporary storage the same amount of data in the same cases as hash-based aggregation. For large outputs, wide merging uses an in-memory index during the final merge step in order to combine many more runs than a traditional merge step. These two new techniques ensure that sort-based duplicate removal, grouping, and aggregation is always competitive with hash-based query execution; it is superior if interesting orderings matter.

In conclusion, a single algorithm for duplicate removal, grouping, and aggregation provides multiple benefits in a query execution engine. Most obviously, it reduces the code volume and maintenance burden for query execution. Perhaps more importantly, it eliminates from query optimization the danger of mistaken algorithm choices (at least for duplicate removal, grouping, and aggregation). It also eliminates unwelcome performance surprises, unhappy users due to unpredictable algorithm choices, and engineering time wasted on analyzing execution traces. Predictable performance, in particular when combined with graceful degradation, permits more smoothly-running applications, more responsive dashboards, and more productive users.

\begin{acks}
We appreciate all the encouragement and help we received from John Cieslewicz, Jagan Sankaranarayanan, and the entire F1~Query and Napa teams at Google.
\end{acks}

\bibliographystyle{ACM-Reference-Format}
\bibliography{in_sort_agg}


\begin{thebibliography}{40}


\ifx \showCODEN    \undefined \def \showCODEN     #1{\unskip}     \fi
\ifx \showDOI      \undefined \def \showDOI       #1{#1}\fi
\ifx \showISBNx    \undefined \def \showISBNx     #1{\unskip}     \fi
\ifx \showISBNxiii \undefined \def \showISBNxiii  #1{\unskip}     \fi
\ifx \showISSN     \undefined \def \showISSN      #1{\unskip}     \fi
\ifx \showLCCN     \undefined \def \showLCCN      #1{\unskip}     \fi
\ifx \shownote     \undefined \def \shownote      #1{#1}          \fi
\ifx \showarticletitle \undefined \def \showarticletitle #1{#1}   \fi
\ifx \showURL      \undefined \def \showURL       {\relax}        \fi
\providecommand\bibfield[2]{#2}
\providecommand\bibinfo[2]{#2}
\providecommand\natexlab[1]{#1}
\providecommand\showeprint[2][]{arXiv:#2}

\bibitem[\protect\citeauthoryear{??}{TPC}{2021}]%
        {TPC}
 \bibinfo{year}{2021}\natexlab{}.
\newblock \bibinfo{title}{{TPC-H} benchmark}.
\newblock
\newblock
\urldef\tempurl%
\url{http://www.tpc.org/tpch/}
\showURL{%
\tempurl}


\bibitem[\protect\citeauthoryear{Agiwal, Lai, Manoharan, Roy, Sankaranarayanan,
  Zhang, Zou, Chen, Chen, Dai, Do, Gao, Geng, Grover, Huang, Huang, Li, Liang,
  Lin, Liu, Liu, Mao, Meng, Mishra, Patel, R, Raman, Roy, Shishodia, Sun, Tang,
  Tatemura, Trehan, Vadali, Venkatasubramanian, Zhang, Zhang, Zhang, Zhuang,
  Graefe, Agrawal, Naughton, Kosalge, and Hacıgümüş}{Agiwal
  et~al\mbox{.}}{2021}]%
        {Napa21}
\bibfield{author}{\bibinfo{person}{Ankur Agiwal}, \bibinfo{person}{Kevin Lai},
  \bibinfo{person}{Gokul Nath~Babu Manoharan}, \bibinfo{person}{Indrajit Roy},
  \bibinfo{person}{Jagan Sankaranarayanan}, \bibinfo{person}{Hao Zhang},
  \bibinfo{person}{Tao Zou}, \bibinfo{person}{Min Chen}, \bibinfo{person}{Jim
  Chen}, \bibinfo{person}{Ming Dai}, \bibinfo{person}{Thanh Do},
  \bibinfo{person}{Haoyu Gao}, \bibinfo{person}{Haoyan Geng},
  \bibinfo{person}{Raman Grover}, \bibinfo{person}{Bo Huang},
  \bibinfo{person}{Yanlai Huang}, \bibinfo{person}{Adam Li},
  \bibinfo{person}{Jianyi Liang}, \bibinfo{person}{Tao Lin},
  \bibinfo{person}{Li Liu}, \bibinfo{person}{Yao Liu}, \bibinfo{person}{Xi
  Mao}, \bibinfo{person}{Maya Meng}, \bibinfo{person}{Prashant Mishra},
  \bibinfo{person}{Jay Patel}, \bibinfo{person}{Rajesh~S R},
  \bibinfo{person}{Vijayshankar Raman}, \bibinfo{person}{Sourashis Roy},
  \bibinfo{person}{Mayank~Singh Shishodia}, \bibinfo{person}{Tianhang Sun},
  \bibinfo{person}{Justin Tang}, \bibinfo{person}{Junichi Tatemura},
  \bibinfo{person}{Sagar Trehan}, \bibinfo{person}{Ramkumar Vadali},
  \bibinfo{person}{Prasanna Venkatasubramanian}, \bibinfo{person}{Joey Zhang},
  \bibinfo{person}{Kefei Zhang}, \bibinfo{person}{Yupu Zhang},
  \bibinfo{person}{Zeleng Zhuang}, \bibinfo{person}{Goetz Graefe},
  \bibinfo{person}{Divyakanth Agrawal}, \bibinfo{person}{Jeff Naughton},
  \bibinfo{person}{Sujata~Sunil Kosalge}, {and} \bibinfo{person}{Hakan
  Hacıgümüş}.} \bibinfo{year}{2021}\natexlab{}.
\newblock \showarticletitle{Napa: Powering Scalable Data Warehousing with
  Robust Query Performance at Google}.
\newblock \bibinfo{journal}{\emph{Proceedings of the VLDB Endowment (PVLDB)}}
  \bibinfo{volume}{14 (12)} (\bibinfo{year}{2021}),
  \bibinfo{pages}{2986--2998}.
\newblock


\bibitem[\protect\citeauthoryear{Bayer and McCreight}{Bayer and
  McCreight}{1972}]%
        {BM:72}
\bibfield{author}{\bibinfo{person}{R. Bayer} {and} \bibinfo{person}{E.
  McCreight}.} \bibinfo{year}{1972}\natexlab{}.
\newblock \showarticletitle{Organization and Maintenance of Large Ordered
  Indices}.
\newblock \bibinfo{journal}{\emph{Acta Informatica}}  \bibinfo{volume}{1}
  (\bibinfo{year}{1972}), \bibinfo{pages}{173–189}.
\newblock
\urldef\tempurl%
\url{https://doi.org/10.1007/BF00288683}
\showDOI{\tempurl}


\bibitem[\protect\citeauthoryear{Bitton and DeWitt}{Bitton and DeWitt}{1983}]%
        {BD:83}
\bibfield{author}{\bibinfo{person}{Dina Bitton} {and} \bibinfo{person}{David~J.
  DeWitt}.} \bibinfo{year}{1983}\natexlab{}.
\newblock \showarticletitle{Duplicate Record Elimination in Large Data Files}.
\newblock \bibinfo{journal}{\emph{ACM Trans. Database Syst.}}
  \bibinfo{volume}{8}, \bibinfo{number}{2} (\bibinfo{date}{June}
  \bibinfo{year}{1983}), \bibinfo{pages}{255–265}.
\newblock
\showISSN{0362-5915}
\urldef\tempurl%
\url{https://doi.org/10.1145/319983.319987}
\showDOI{\tempurl}


\bibitem[\protect\citeauthoryear{Boncz, Neumann, and Erling}{Boncz
  et~al\mbox{.}}{2013}]%
        {BNE:13}
\bibfield{author}{\bibinfo{person}{Peter Boncz}, \bibinfo{person}{Thomas
  Neumann}, {and} \bibinfo{person}{Orri Erling}.}
  \bibinfo{year}{2013}\natexlab{}.
\newblock \showarticletitle{{TPC-H} Analyzed: {H}idden Messages and Lessons
  Learned from an Influential Benchmark}. In \bibinfo{booktitle}{\emph{Revised
  Selected Papers of the 5th TPC Technology Conference on Performance
  Characterization and Benchmarking - Volume 8391}}.
  \bibinfo{publisher}{Springer-Verlag}, \bibinfo{address}{Berlin, Heidelberg},
  \bibinfo{pages}{61–76}.
\newblock
\showISBNx{9783319049359}
\urldef\tempurl%
\url{https://doi.org/10.1007/978-3-319-04936-6_5}
\showDOI{\tempurl}


\bibitem[\protect\citeauthoryear{Chang, Dean, Ghemawat, Hsieh, Wallach,
  Burrows, Chandra, Fikes, and Gruber}{Chang et~al\mbox{.}}{2008}]%
        {C+:08}
\bibfield{author}{\bibinfo{person}{Fay Chang}, \bibinfo{person}{Jeffrey Dean},
  \bibinfo{person}{Sanjay Ghemawat}, \bibinfo{person}{Wilson~C. Hsieh},
  \bibinfo{person}{Deborah~A. Wallach}, \bibinfo{person}{Mike Burrows},
  \bibinfo{person}{Tushar Chandra}, \bibinfo{person}{Andrew Fikes}, {and}
  \bibinfo{person}{Robert~E. Gruber}.} \bibinfo{year}{2008}\natexlab{}.
\newblock \showarticletitle{Big{T}able: {A} {D}istributed {S}torage {S}ystem
  for {S}tructured {D}ata}.
\newblock \bibinfo{journal}{\emph{ACM Trans. Comput. Syst.}}
  \bibinfo{volume}{26}, \bibinfo{number}{2}, Article \bibinfo{articleno}{4}
  (\bibinfo{date}{June} \bibinfo{year}{2008}), \bibinfo{numpages}{26}~pages.
\newblock
\showISSN{0734-2071}
\urldef\tempurl%
\url{https://doi.org/10.1145/1365815.1365816}
\showDOI{\tempurl}


\bibitem[\protect\citeauthoryear{Conner}{Conner}{1977}]%
        {C:77}
\bibfield{author}{\bibinfo{person}{W.~M. Conner}.}
  \bibinfo{year}{1977}\natexlab{}.
\newblock \showarticletitle{Offset-value coding}. In
  \bibinfo{booktitle}{\emph{IBM Technical Disclosure Bulletin}}.
\newblock


\bibitem[\protect\citeauthoryear{Corbett, Dean, Epstein, Fikes, Frost, Furman,
  Ghemawat, Gubarev, Heiser, Hochschild, Hsieh, Kanthak, Kogan, Li, Lloyd,
  Melnik, Mwaura, Nagle, Quinlan, Rao, Rolig, Saito, Szymaniak, Taylor, Wang,
  and Woodford}{Corbett et~al\mbox{.}}{2013}]%
        {C+:13}
\bibfield{author}{\bibinfo{person}{James~C. Corbett}, \bibinfo{person}{Jeffrey
  Dean}, \bibinfo{person}{Michael Epstein}, \bibinfo{person}{Andrew Fikes},
  \bibinfo{person}{Christopher Frost}, \bibinfo{person}{J.~J. Furman},
  \bibinfo{person}{Sanjay Ghemawat}, \bibinfo{person}{Andrey Gubarev},
  \bibinfo{person}{Christopher Heiser}, \bibinfo{person}{Peter Hochschild},
  \bibinfo{person}{Wilson Hsieh}, \bibinfo{person}{Sebastian Kanthak},
  \bibinfo{person}{Eugene Kogan}, \bibinfo{person}{Hongyi Li},
  \bibinfo{person}{Alexander Lloyd}, \bibinfo{person}{Sergey Melnik},
  \bibinfo{person}{David Mwaura}, \bibinfo{person}{David Nagle},
  \bibinfo{person}{Sean Quinlan}, \bibinfo{person}{Rajesh Rao},
  \bibinfo{person}{Lindsay Rolig}, \bibinfo{person}{Yasushi Saito},
  \bibinfo{person}{Michal Szymaniak}, \bibinfo{person}{Christopher Taylor},
  \bibinfo{person}{Ruth Wang}, {and} \bibinfo{person}{Dale Woodford}.}
  \bibinfo{year}{2013}\natexlab{}.
\newblock \showarticletitle{{S}panner: {G}oogle’s {G}lobally {D}istributed
  {D}atabase}.
\newblock \bibinfo{journal}{\emph{ACM Trans. Comput. Syst.}}
  \bibinfo{volume}{31}, \bibinfo{number}{3}, Article \bibinfo{articleno}{8}
  (\bibinfo{date}{Aug.} \bibinfo{year}{2013}), \bibinfo{numpages}{22}~pages.
\newblock
\showISSN{0734-2071}
\urldef\tempurl%
\url{https://doi.org/10.1145/2491245}
\showDOI{\tempurl}


\bibitem[\protect\citeauthoryear{Dean and Ghemawat}{Dean and Ghemawat}{2008}]%
        {DG:08}
\bibfield{author}{\bibinfo{person}{Jeffrey Dean} {and} \bibinfo{person}{Sanjay
  Ghemawat}.} \bibinfo{year}{2008}\natexlab{}.
\newblock \showarticletitle{{M}ap{R}educe: {S}implified {D}ata {P}rocessing on
  {L}arge {C}lusters}.
\newblock \bibinfo{journal}{\emph{Commun. ACM}} \bibinfo{volume}{51},
  \bibinfo{number}{1} (\bibinfo{date}{Jan.} \bibinfo{year}{2008}),
  \bibinfo{pages}{107–113}.
\newblock
\showISSN{0001-0782}
\urldef\tempurl%
\url{https://doi.org/10.1145/1327452.1327492}
\showDOI{\tempurl}


\bibitem[\protect\citeauthoryear{Dittrich, Seeger, Taylor, and
  Widmayer}{Dittrich et~al\mbox{.}}{2003}]%
        {DST:03}
\bibfield{author}{\bibinfo{person}{Jens-Peter Dittrich},
  \bibinfo{person}{Bernhard Seeger}, \bibinfo{person}{David~Scot Taylor}, {and}
  \bibinfo{person}{Peter Widmayer}.} \bibinfo{year}{2003}\natexlab{}.
\newblock \showarticletitle{On Producing Join Results Early}. In
  \bibinfo{booktitle}{\emph{ACM PODS'03}}. \bibinfo{pages}{134–142}.
\newblock
\showISBNx{1581136706}
\urldef\tempurl%
\url{https://doi.org/10.1145/773153.773167}
\showDOI{\tempurl}


\bibitem[\protect\citeauthoryear{Epstein}{Epstein}{1979}]%
        {E:79}
\bibfield{author}{\bibinfo{person}{Robert Epstein}.}
  \bibinfo{year}{1979}\natexlab{}.
\newblock \showarticletitle{Techniques for processing of aggregates in
  relational database systems}. In \bibinfo{booktitle}{\emph{Univ. of
  California at Berkeley, UCB/ERL Memorandum M79/8}}.
\newblock


\bibitem[\protect\citeauthoryear{Gassner}{Gassner}{1967}]%
        {G:67}
\bibfield{author}{\bibinfo{person}{Betty~Jane Gassner}.}
  \bibinfo{year}{1967}\natexlab{}.
\newblock \showarticletitle{Sorting by Replacement Selecting}.
\newblock \bibinfo{journal}{\emph{Commun. ACM}} \bibinfo{volume}{10},
  \bibinfo{number}{2} (\bibinfo{date}{Feb.} \bibinfo{year}{1967}),
  \bibinfo{pages}{89–93}.
\newblock
\showISSN{0001-0782}
\urldef\tempurl%
\url{https://doi.org/10.1145/363067.363102}
\showDOI{\tempurl}


\bibitem[\protect\citeauthoryear{Goetz}{Goetz}{1963}]%
        {G:63}
\bibfield{author}{\bibinfo{person}{Martin~A. Goetz}.}
  \bibinfo{year}{1963}\natexlab{}.
\newblock \showarticletitle{Internal and Tape Sorting Using the
  Replacement-Selection Technique}.
\newblock \bibinfo{journal}{\emph{Commun. ACM}} \bibinfo{volume}{6},
  \bibinfo{number}{5} (\bibinfo{date}{May} \bibinfo{year}{1963}),
  \bibinfo{pages}{201–206}.
\newblock
\showISSN{0001-0782}
\urldef\tempurl%
\url{https://doi.org/10.1145/366552.366556}
\showDOI{\tempurl}


\bibitem[\protect\citeauthoryear{Graefe}{Graefe}{1993}]%
        {G:93}
\bibfield{author}{\bibinfo{person}{Goetz Graefe}.}
  \bibinfo{year}{1993}\natexlab{}.
\newblock \showarticletitle{Query {E}valuation {T}echniques for {L}arge
  {D}atabases}.
\newblock \bibinfo{journal}{\emph{ACM Comput. Surv.}} \bibinfo{volume}{25},
  \bibinfo{number}{2} (\bibinfo{date}{June} \bibinfo{year}{1993}),
  \bibinfo{pages}{73--169}.
\newblock
\showISSN{0360-0300}
\urldef\tempurl%
\url{https://doi.org/10.1145/152610.152611}
\showDOI{\tempurl}


\bibitem[\protect\citeauthoryear{Graefe}{Graefe}{1994}]%
        {G:94}
\bibfield{author}{\bibinfo{person}{G. Graefe}.}
  \bibinfo{year}{1994}\natexlab{}.
\newblock \showarticletitle{Volcano— {A}n {E}xtensible and {P}arallel {Q}uery
  {E}valuation {S}ystem}.
\newblock \bibinfo{journal}{\emph{IEEE Trans. on Knowl. and Data Eng.}}
  \bibinfo{volume}{6}, \bibinfo{number}{1} (\bibinfo{date}{Feb.}
  \bibinfo{year}{1994}), \bibinfo{pages}{120–135}.
\newblock
\showISSN{1041-4347}
\urldef\tempurl%
\url{https://doi.org/10.1109/69.273032}
\showDOI{\tempurl}


\bibitem[\protect\citeauthoryear{Graefe}{Graefe}{2006}]%
        {G:06}
\bibfield{author}{\bibinfo{person}{Goetz Graefe}.}
  \bibinfo{year}{2006}\natexlab{}.
\newblock \showarticletitle{Implementing {S}orting in {D}atabase {S}ystems}.
\newblock \bibinfo{journal}{\emph{ACM Comput. Surv.}} \bibinfo{volume}{38},
  \bibinfo{number}{3}, Article \bibinfo{articleno}{10} (\bibinfo{date}{Sept.}
  \bibinfo{year}{2006}).
\newblock
\showISSN{0360-0300}
\urldef\tempurl%
\url{https://doi.org/10.1145/1132960.1132964}
\showDOI{\tempurl}


\bibitem[\protect\citeauthoryear{Graefe}{Graefe}{2011a}]%
        {G:11:gjoin}
\bibfield{author}{\bibinfo{person}{Goetz Graefe}.}
  \bibinfo{year}{2011}\natexlab{a}.
\newblock \showarticletitle{A Generalized Join Algorithm}. In
  \bibinfo{booktitle}{\emph{Datenbanksysteme f{\"{u}}r Business, Technologie
  und Web (BTW), 14. Fachtagung des GI-Fachbereichs "Datenbanken und
  Informationssysteme" (DBIS), 2.-4.3.2011 in Kaiserslautern, Germany}}
  \emph{(\bibinfo{series}{{LNI}}, Vol.~\bibinfo{volume}{{P-180}})},
  \bibfield{editor}{\bibinfo{person}{Theo H{\"{a}}rder},
  \bibinfo{person}{Wolfgang Lehner}, \bibinfo{person}{Bernhard Mitschang},
  \bibinfo{person}{Harald Sch{\"{o}}ning}, {and} \bibinfo{person}{Holger
  Schwarz}} (Eds.). \bibinfo{publisher}{{GI}}, \bibinfo{pages}{267--286}.
\newblock
\urldef\tempurl%
\url{https://dl.gi.de/20.500.12116/19583}
\showURL{%
\tempurl}


\bibitem[\protect\citeauthoryear{Graefe}{Graefe}{2011b}]%
        {G:11}
\bibfield{author}{\bibinfo{person}{Goetz Graefe}.}
  \bibinfo{year}{2011}\natexlab{b}.
\newblock \showarticletitle{Modern {B}-{T}ree {T}echniques}.
\newblock \bibinfo{journal}{\emph{Found. Trends Databases}}
  \bibinfo{volume}{3}, \bibinfo{number}{4} (\bibinfo{date}{April}
  \bibinfo{year}{2011}), \bibinfo{pages}{203--402}.
\newblock
\showISSN{1931-7883}
\urldef\tempurl%
\url{https://doi.org/10.1561/1900000028}
\showDOI{\tempurl}


\bibitem[\protect\citeauthoryear{Graefe}{Graefe}{2012}]%
        {G:12:gjoin}
\bibfield{author}{\bibinfo{person}{Goetz Graefe}.}
  \bibinfo{year}{2012}\natexlab{}.
\newblock \showarticletitle{New algorithms for join and grouping operations}.
\newblock \bibinfo{journal}{\emph{Comput. Sci. Res. Dev.}}
  \bibinfo{volume}{27}, \bibinfo{number}{1} (\bibinfo{year}{2012}),
  \bibinfo{pages}{3--27}.
\newblock
\urldef\tempurl%
\url{https://doi.org/10.1007/s00450-011-0186-9}
\showDOI{\tempurl}


\bibitem[\protect\citeauthoryear{Graefe, Bunker, and Cooper}{Graefe
  et~al\mbox{.}}{1998}]%
        {GBC:98}
\bibfield{author}{\bibinfo{person}{Goetz Graefe}, \bibinfo{person}{Ross
  Bunker}, {and} \bibinfo{person}{Shaun Cooper}.}
  \bibinfo{year}{1998}\natexlab{}.
\newblock \showarticletitle{{H}ash {J}oins and {H}ash {T}eams in {M}icrosoft
  {SQL} {S}erver}. In \bibinfo{booktitle}{\emph{VLDB '98}}.
  \bibinfo{pages}{86–97}.
\newblock
\showISBNx{1558605665}


\bibitem[\protect\citeauthoryear{Graefe, Guy, and Sauer}{Graefe
  et~al\mbox{.}}{2016}]%
        {GGS:16}
\bibfield{author}{\bibinfo{person}{Goetz Graefe}, \bibinfo{person}{Wey Guy},
  {and} \bibinfo{person}{Caetano Sauer}.} \bibinfo{year}{2016}\natexlab{}.
\newblock \showarticletitle{Instant recovery with write-ahead logging: page
  repair, system restart, media restore, and system failover}.
\newblock \bibinfo{journal}{\emph{Synthesis Lectures on Data Management}}
  \bibinfo{volume}{8}, \bibinfo{number}{2} (\bibinfo{year}{2016}),
  \bibinfo{pages}{1--113}.
\newblock


\bibitem[\protect\citeauthoryear{Gray}{Gray}{1978}]%
        {G:78}
\bibfield{author}{\bibinfo{person}{Jim Gray}.} \bibinfo{year}{1978}\natexlab{}.
\newblock \showarticletitle{Notes on {D}atabase {O}perating {S}ystems}. In
  \bibinfo{booktitle}{\emph{Operating Systems, An Advanced Course}}.
  \bibinfo{publisher}{Springer-Verlag}, \bibinfo{address}{Berlin, Heidelberg},
  \bibinfo{pages}{393–481}.
\newblock
\showISBNx{3540087559}


\bibitem[\protect\citeauthoryear{Gupta, Yang, Govig, Kirsch, Chan, Lai, Wu,
  Dhoot, Kumar, Agiwal, Bhansali, Hong, Cameron, Siddiqi, Jones, Shute,
  Gubarev, Venkataraman, and Agrawal}{Gupta et~al\mbox{.}}{2016}]%
        {G+:16}
\bibfield{author}{\bibinfo{person}{Ashish Gupta}, \bibinfo{person}{Fan Yang},
  \bibinfo{person}{Jason Govig}, \bibinfo{person}{Adam Kirsch},
  \bibinfo{person}{Kelvin Chan}, \bibinfo{person}{Kevin Lai},
  \bibinfo{person}{Shuo Wu}, \bibinfo{person}{Sandeep Dhoot},
  \bibinfo{person}{Abhilash~Rajesh Kumar}, \bibinfo{person}{Ankur Agiwal},
  \bibinfo{person}{Sanjay Bhansali}, \bibinfo{person}{Mingsheng Hong},
  \bibinfo{person}{Jamie Cameron}, \bibinfo{person}{Masood Siddiqi},
  \bibinfo{person}{David Jones}, \bibinfo{person}{Jeff Shute},
  \bibinfo{person}{Andrey Gubarev}, \bibinfo{person}{Shivakumar Venkataraman},
  {and} \bibinfo{person}{Divyakant Agrawal}.} \bibinfo{year}{2016}\natexlab{}.
\newblock \showarticletitle{Mesa: {A} {G}eo-{R}eplicated {O}nline {D}ata
  {W}arehouse for {G}oogle's {A}dvertising {S}ystem}.
\newblock \bibinfo{journal}{\emph{Commun. ACM}} \bibinfo{volume}{59},
  \bibinfo{number}{7} (\bibinfo{date}{June} \bibinfo{year}{2016}),
  \bibinfo{pages}{117–125}.
\newblock
\showISSN{0001-0782}
\urldef\tempurl%
\url{https://doi.org/10.1145/2936722}
\showDOI{\tempurl}


\bibitem[\protect\citeauthoryear{H\"{a}rder}{H\"{a}rder}{1977}]%
        {H:77}
\bibfield{author}{\bibinfo{person}{Theo H\"{a}rder}.}
  \bibinfo{year}{1977}\natexlab{}.
\newblock \showarticletitle{A {S}can-{D}riven {S}ort {F}acility for a
  {R}elational {D}atabase {S}ystem}. In \bibinfo{booktitle}{\emph{VLDB '77}}
  (Tokyo, Japan). \bibinfo{pages}{236–244}.
\newblock


\bibitem[\protect\citeauthoryear{Hellerstein and Naughton}{Hellerstein and
  Naughton}{1996}]%
        {HN:96}
\bibfield{author}{\bibinfo{person}{Joseph~M. Hellerstein} {and}
  \bibinfo{person}{Jeffrey~F. Naughton}.} \bibinfo{year}{1996}\natexlab{}.
\newblock \showarticletitle{Query {E}xecution {T}echniques for {C}aching
  {E}xpensive {M}ethods}. In \bibinfo{booktitle}{\emph{ACM SIGMOD '96}}.
  \bibinfo{pages}{423–434}.
\newblock
\showISBNx{0897917944}
\urldef\tempurl%
\url{https://doi.org/10.1145/233269.233359}
\showDOI{\tempurl}


\bibitem[\protect\citeauthoryear{Iyer}{Iyer}{2005}]%
        {I:05}
\bibfield{author}{\bibinfo{person}{Bala~R. Iyer}.}
  \bibinfo{year}{2005}\natexlab{}.
\newblock \showarticletitle{Hardware Assisted Sorting in {IBM}'s {DB2} {DBMS}}.
  In \bibinfo{booktitle}{\emph{International Conference on Management of Data
  (COMAD)}}.
\newblock


\bibitem[\protect\citeauthoryear{Jagadish, Narayan, Seshadri, Sudarshan, and
  Kanneganti}{Jagadish et~al\mbox{.}}{[n.d.]}]%
        {JNS:97}
\bibfield{author}{\bibinfo{person}{H.~V. Jagadish}, \bibinfo{person}{P.~P.~S.
  Narayan}, \bibinfo{person}{S. Seshadri}, \bibinfo{person}{S. Sudarshan},
  {and} \bibinfo{person}{Rama Kanneganti}.} \bibinfo{year}{[n.d.]}\natexlab{}.
\newblock \showarticletitle{Incremental {O}rganization for {D}ata {R}ecording
  and {W}arehousing}. In \bibinfo{booktitle}{\emph{VLDB '97}}.
  \bibinfo{pages}{16–25}.
\newblock
\showISBNx{1558604707}


\bibitem[\protect\citeauthoryear{Knuth}{Knuth}{1998}]%
        {K:98}
\bibfield{author}{\bibinfo{person}{Donald~E. Knuth}.}
  \bibinfo{year}{1998}\natexlab{}.
\newblock \bibinfo{booktitle}{\emph{The Art of Computer Programming, Volume 3:
  (2nd Ed.) Sorting and Searching}}.
\newblock \bibinfo{publisher}{Addison Wesley Longman Publishing Co., Inc.},
  \bibinfo{address}{USA}.
\newblock
\showISBNx{0201896850}


\bibitem[\protect\citeauthoryear{Kooi}{Kooi}{1980}]%
        {K:80}
\bibfield{author}{\bibinfo{person}{Robert~Philip Kooi}.}
  \bibinfo{year}{1980}\natexlab{}.
\newblock \emph{\bibinfo{title}{The Optimization of Queries in Relational
  Databases}}.
\newblock \bibinfo{thesistype}{Ph.D. Dissertation}. \bibinfo{address}{USA}.
\newblock


\bibitem[\protect\citeauthoryear{M\"{u}ller, Sanders, Lacurie, Lehner, and
  F\"{a}rber}{M\"{u}ller et~al\mbox{.}}{2015}]%
        {MSL:15}
\bibfield{author}{\bibinfo{person}{Ingo M\"{u}ller}, \bibinfo{person}{Peter
  Sanders}, \bibinfo{person}{Arnaud Lacurie}, \bibinfo{person}{Wolfgang
  Lehner}, {and} \bibinfo{person}{Franz F\"{a}rber}.}
  \bibinfo{year}{2015}\natexlab{}.
\newblock \showarticletitle{Cache-{E}fficient {A}ggregation: {H}ashing {I}s
  {S}orting}. In \bibinfo{booktitle}{\emph{ACM SIGMOD '15}}.
  \bibinfo{pages}{1123–1136}.
\newblock
\showISBNx{9781450327589}
\urldef\tempurl%
\url{https://doi.org/10.1145/2723372.2747644}
\showDOI{\tempurl}


\bibitem[\protect\citeauthoryear{Neumann and Moerkotte}{Neumann and
  Moerkotte}{2004}]%
        {NM:04}
\bibfield{author}{\bibinfo{person}{Thomas Neumann} {and} \bibinfo{person}{Guido
  Moerkotte}.} \bibinfo{year}{2004}\natexlab{}.
\newblock \showarticletitle{A {C}ombined {F}ramework for {G}rouping and {O}rder
  {O}ptimization}. In \bibinfo{booktitle}{\emph{VLDB '04}}.
  \bibinfo{pages}{960–971}.
\newblock
\showISBNx{0120884690}


\bibitem[\protect\citeauthoryear{O'Neil, Cheng, Gawlick, and O'Neil}{O'Neil
  et~al\mbox{.}}{1996}]%
        {OCG:96}
\bibfield{author}{\bibinfo{person}{Patrick O'Neil}, \bibinfo{person}{Edward
  Cheng}, \bibinfo{person}{Dieter Gawlick}, {and} \bibinfo{person}{Elizabeth
  O'Neil}.} \bibinfo{year}{1996}\natexlab{}.
\newblock \showarticletitle{The {L}og-structured {M}erge-tree ({LSM}-tree)}.
\newblock \bibinfo{journal}{\emph{Acta Informatica}} \bibinfo{volume}{33},
  \bibinfo{number}{4} (\bibinfo{date}{June} \bibinfo{year}{1996}),
  \bibinfo{pages}{351--385}.
\newblock
\showISSN{0001-5903}
\urldef\tempurl%
\url{https://doi.org/10.1007/s002360050048}
\showDOI{\tempurl}


\bibitem[\protect\citeauthoryear{Pang, Carey, and Livny}{Pang
  et~al\mbox{.}}{1993}]%
        {PangCareyLivny-93-Sorting}
\bibfield{author}{\bibinfo{person}{HweeHwa Pang}, \bibinfo{person}{Michael~J.
  Carey}, {and} \bibinfo{person}{Miron Livny}.}
  \bibinfo{year}{1993}\natexlab{}.
\newblock \showarticletitle{Memory-Adaptive External Sorting}. In
  \bibinfo{booktitle}{\emph{19th International Conference on Very Large Data
  Bases, August 24-27, 1993, Dublin, Ireland}},
  \bibfield{editor}{\bibinfo{person}{Rakesh Agrawal},
  \bibinfo{person}{Se{\'{a}}n Baker}, {and} \bibinfo{person}{David~A. Bell}}
  (Eds.). \bibinfo{publisher}{Morgan Kaufmann}, \bibinfo{pages}{618--629}.
\newblock


\bibitem[\protect\citeauthoryear{Samwel, Cieslewicz, Handy, Govig, Venetis,
  Yang, Peters, Shute, Tenedorio, Apte, Weigel, Wilhite, Yang, Xu, Li, Yuan,
  Chasseur, Zeng, Rae, Biyani, Harn, Xia, Gubichev, El-Helw, Erling, Yan, Yang,
  Wei, Do, Zheng, Graefe, Sardashti, Aly, Agrawal, Gupta, and
  Venkataraman}{Samwel et~al\mbox{.}}{2018}]%
        {S+:18}
\bibfield{author}{\bibinfo{person}{Bart Samwel}, \bibinfo{person}{John
  Cieslewicz}, \bibinfo{person}{Ben Handy}, \bibinfo{person}{Jason Govig},
  \bibinfo{person}{Petros Venetis}, \bibinfo{person}{Chanjun Yang},
  \bibinfo{person}{Keith Peters}, \bibinfo{person}{Jeff Shute},
  \bibinfo{person}{Daniel Tenedorio}, \bibinfo{person}{Himani Apte},
  \bibinfo{person}{Felix Weigel}, \bibinfo{person}{David Wilhite},
  \bibinfo{person}{Jiacheng Yang}, \bibinfo{person}{Jun Xu},
  \bibinfo{person}{Jiexing Li}, \bibinfo{person}{Zhan Yuan},
  \bibinfo{person}{Craig Chasseur}, \bibinfo{person}{Qiang Zeng},
  \bibinfo{person}{Ian Rae}, \bibinfo{person}{Anurag Biyani},
  \bibinfo{person}{Andrew Harn}, \bibinfo{person}{Yang Xia},
  \bibinfo{person}{Andrey Gubichev}, \bibinfo{person}{Amr El-Helw},
  \bibinfo{person}{Orri Erling}, \bibinfo{person}{Zhepeng Yan},
  \bibinfo{person}{Mohan Yang}, \bibinfo{person}{Yiqun Wei},
  \bibinfo{person}{Thanh Do}, \bibinfo{person}{Colin Zheng},
  \bibinfo{person}{Goetz Graefe}, \bibinfo{person}{Somayeh Sardashti},
  \bibinfo{person}{Ahmed~M. Aly}, \bibinfo{person}{Divy Agrawal},
  \bibinfo{person}{Ashish Gupta}, {and} \bibinfo{person}{Shiv Venkataraman}.}
  \bibinfo{year}{2018}\natexlab{}.
\newblock \showarticletitle{F1 {Q}uery: {D}eclarative {Q}uerying at {S}cale}.
  In \bibinfo{booktitle}{\emph{VLDB '18}}. \bibinfo{pages}{1835--1848}.
\newblock
\showISSN{2150-8097}
\urldef\tempurl%
\url{https://doi.org/10.14778/3229863.3229871}
\showDOI{\tempurl}


\bibitem[\protect\citeauthoryear{Selinger, Astrahan, Chamberlin, Lorie, and
  Price}{Selinger et~al\mbox{.}}{1979}]%
        {SAC:79}
\bibfield{author}{\bibinfo{person}{P.~Griffiths Selinger},
  \bibinfo{person}{M.~M. Astrahan}, \bibinfo{person}{D.~D. Chamberlin},
  \bibinfo{person}{R.~A. Lorie}, {and} \bibinfo{person}{T.~G. Price}.}
  \bibinfo{year}{1979}\natexlab{}.
\newblock \showarticletitle{Access {P}ath {S}election in a {R}elational
  {D}atabase {M}anagement {S}ystem}. In \bibinfo{booktitle}{\emph{ACM SIGMOD
  '79}}. \bibinfo{pages}{23–34}.
\newblock
\showISBNx{089791001X}
\urldef\tempurl%
\url{https://doi.org/10.1145/582095.582099}
\showDOI{\tempurl}


\bibitem[\protect\citeauthoryear{Shapiro}{Shapiro}{1986}]%
        {Shapiro-86-Hybrid-hash}
\bibfield{author}{\bibinfo{person}{Leonard~D. Shapiro}.}
  \bibinfo{year}{1986}\natexlab{}.
\newblock \showarticletitle{Join Processing in Database Systems with Large Main
  Memories}.
\newblock \bibinfo{journal}{\emph{ACM Transactions on Database Systems}}
  \bibinfo{volume}{11}, \bibinfo{number}{3} (\bibinfo{date}{Aug.}
  \bibinfo{year}{1986}), \bibinfo{pages}{239–264}.
\newblock
\showISSN{0362-5915}
\urldef\tempurl%
\url{https://doi.org/10.1145/6314.6315}
\showDOI{\tempurl}


\bibitem[\protect\citeauthoryear{Shute, Vingralek, Samwel, Handy, Whipkey,
  Rollins, Oancea, Littlefield, Menestrina, Ellner, Cieslewicz, Rae, Stancescu,
  and Apte}{Shute et~al\mbox{.}}{2013}]%
        {S+:13}
\bibfield{author}{\bibinfo{person}{Jeff Shute}, \bibinfo{person}{Radek
  Vingralek}, \bibinfo{person}{Bart Samwel}, \bibinfo{person}{Ben Handy},
  \bibinfo{person}{Chad Whipkey}, \bibinfo{person}{Eric Rollins},
  \bibinfo{person}{Mircea Oancea}, \bibinfo{person}{Kyle Littlefield},
  \bibinfo{person}{David Menestrina}, \bibinfo{person}{Stephan Ellner},
  \bibinfo{person}{John Cieslewicz}, \bibinfo{person}{Ian Rae},
  \bibinfo{person}{Traian Stancescu}, {and} \bibinfo{person}{Himani Apte}.}
  \bibinfo{year}{2013}\natexlab{}.
\newblock \showarticletitle{F1: {A} {D}istributed {SQL} {D}atabase {T}hat
  {S}cales}. In \bibinfo{booktitle}{\emph{VLDB '13}}.
  \bibinfo{pages}{1068–1079}.
\newblock
\showISSN{2150-8097}
\urldef\tempurl%
\url{https://doi.org/10.14778/2536222.2536232}
\showDOI{\tempurl}


\bibitem[\protect\citeauthoryear{Simmen, Shekita, and Malkemus}{Simmen
  et~al\mbox{.}}{1996}]%
        {SSM:96}
\bibfield{author}{\bibinfo{person}{David Simmen}, \bibinfo{person}{Eugene
  Shekita}, {and} \bibinfo{person}{Timothy Malkemus}.}
  \bibinfo{year}{1996}\natexlab{}.
\newblock \showarticletitle{Fundamental {T}echniques for {O}rder
  {O}ptimization}. In \bibinfo{booktitle}{\emph{ACM SIGMOD '96}} (Montreal,
  Quebec, Canada). \bibinfo{pages}{57–67}.
\newblock
\showISBNx{0897917944}
\urldef\tempurl%
\url{https://doi.org/10.1145/233269.233320}
\showDOI{\tempurl}


\bibitem[\protect\citeauthoryear{Wilschut and Apers}{Wilschut and
  Apers}{1991}]%
        {WA:91}
\bibfield{author}{\bibinfo{person}{Annita~N. Wilschut} {and}
  \bibinfo{person}{Peter M.~G. Apers}.} \bibinfo{year}{1991}\natexlab{}.
\newblock \showarticletitle{Dataflow {Q}uery {E}xecution in a {P}arallel
  {M}ain-{M}emory {E}nvironment}. In \bibinfo{booktitle}{\emph{PDIS '91}}.
  \bibinfo{pages}{68–77}.
\newblock
\showISBNx{0818622954}


\bibitem[\protect\citeauthoryear{Yan and Larson}{Yan and Larson}{1995}]%
        {YanLarson-95-eager-lazy-aggregation-qo}
\bibfield{author}{\bibinfo{person}{Weipeng~P. Yan} {and}
  \bibinfo{person}{Per{-}{\AA}ke Larson}.} \bibinfo{year}{1995}\natexlab{}.
\newblock \showarticletitle{Eager Aggregation and Lazy Aggregation}. In
  \bibinfo{booktitle}{\emph{VLDB'95, Proceedings of 21th International
  Conference on Very Large Data Bases, September 11-15, 1995, Zurich,
  Switzerland}}, \bibfield{editor}{\bibinfo{person}{Umeshwar Dayal},
  \bibinfo{person}{Peter M.~D. Gray}, {and} \bibinfo{person}{Shojiro Nishio}}
  (Eds.). \bibinfo{publisher}{Morgan Kaufmann}, \bibinfo{pages}{345--357}.
\newblock
\urldef\tempurl%
\url{http://www.vldb.org/conf/1995/P345.PDF}
\showURL{%
\tempurl}


\end{thebibliography}

\appendix

\end{document}